\documentclass[%
 reprint,
superscriptaddress,
 amsmath,amssymb,
  aps,
]{revtex4-2}

\usepackage{graphicx}
\usepackage{dcolumn}
\usepackage[normalem]{ulem}
\usepackage{bm}

\usepackage{physics}
\usepackage{xcolor}
\usepackage{hyperref}

\usepackage{siunitx}

\usepackage{amsmath}
\usepackage{amssymb}

\usepackage{soul}

\usepackage{lipsum}
\usepackage{mathtools}

\begin{document}

\title{A quantum model of charge capture and release onto/from deep traps}

\author{Ivan I. Vrubel}
\email{ivanvrubel@ya.ru}
\affiliation{Ioffe Institute, 194021, Saint Petersburg, Russia}

\author{Vasilii Khanin}
\affiliation{Seaborough BV, Amsterdam, the Netherlands}

\author{Markus Suta}
\affiliation{Inorganic Photoactive Materials, Institute of Inorganic and Structural Chemistry, Heinrich Heine University, Universitätsstraße 1, 40225 Düsseldorf, Germany}

\author{Roman G. Polozkov}
\affiliation{Alferov University, Saint Petersburg, Russia}

\author{Evgeniia D. Cherotchenko}
\affiliation{Ioffe Institute, 194021, Saint Petersburg, Russia}

\begin{abstract}
The rapid development of optical technologies and applications revealed the critical role of point defects affecting device performance.  One of the powerful tools to study influence of defects on charge capture and recombination processes is thermoluminescence. The popular models behind thermoluminescence and carrier capture processes are semi-classic though. They offer good qualitative description, but implicitly exclude quantum nature of the accompanying parameters, such as frequency factors and capture cross sections. As a consequence, results obtained for a specific host material cannot be successfully extrapolated to other materials. Thus, the main purpose of our work is to introduce a reliable analytical model that describes non-radiative capture and release of electrons from/to the \textcolor{black}{conduction band}~(CB). The proposed model is governed by Bose-Einstein statistics (for phonon occupation) and Fermi's golden rule (for resonant charge transfer between the trap and the CB).
The constructed model offers a physical interpretation of the capture coefficients and frequency factors, and seamlessly includes the Coulomb neutral/attractive nature of traps.  It connects the frequency factor to the overlap of wavefunctions of the delocalized CB and trap states, and suggests a strong dependence on the density of charge distribution, i.e. the ionicity/covalency of the chemical bonds within the host. Separation of the resonance condition from the accumulation/dissipation of phonons on the site leads to the conclusion that the capture cross-section does not necessarily depend on the trap depth. 
The model is verified by comparison to reported experimental data, showing good agreement. As such, the model generates reliable information about trap states whose exact nature is not completely understood and allows to do materials research in a more systematic way.
\end{abstract}

\maketitle

\section{Introduction}

Controlled presence of point defects is crucial to performance of optical materials in practical applications: On the one hand intentional doping of phosphors, scintillators, light-emitting diodes (LEDs) or lasing materials is a key to their functional design. On the other hand lack of control over defect formation due to contamination and structural disorder leads to deterioration of durability and efficiency of such materials. E.g. in scintillators trace amounts of impurities form traps that compete with primary recombination sites for free charge carriers and bring undesired afterglow\cite{Shiran18,Lucchini16,Khanin20}. Same in optoelectronics, Shockley-Read-Hall (SRH) process related to the point defects limits the LED efficiency, opening the non-radiative leakage channels \cite{landsberg70,stoneham81,henry77}. For interband mid-IR lasers, threshold current densities are also often limited by defect-related Auger recombination \cite{Razeghi99,Fehse02}. Thus the understanding of the mechanisms underlying the influence of point defects on recombination dynamics represents an important practical problem.

One of the methods to investigate how point defects affect charge carrier transport is thermally stimulated luminescence (TSL). Coupled to paramagnetic resonance spectroscopy \cite{Buryi19, Laguta19} and optical spectroscopy\cite{Spaeth12} it allows to determine the point defects that are detrimental to application  and outline the direction to passivating their negative influence\cite{nikl08, Heggen20, Du20}. However, the extrapolation of the results obtained for a specific host to other materials is usually not successful: thermoluminescence mechanisms are regularly described in a semi-classic manner, which offers a satisfying qualitative interpretation, but lacks connectivity and freedom to insert microscopic explicit nature of point defects. The comprehensive overviews of the state of the art can be found in \cite{sunta15, chen11, yukihara11}. 

Quantum mechanical treatment has been also implemented to describe thermoluminescence\cite{bohm85,bohm87} and charge carrier capture dynamics\cite{Lucovsky65, abakumov91, Turiansky21}. Main approach of the works is based on inclusion of electron-phonon coupling (or multi-phonon emission) into Fermi's golden rule\cite{abakumov91, Alkauskas14, Alkauskas16, Wickramaratne18}. As such, Bohm et al. arrived at significantly different temperature dependence of release rates than a semi-classic model\cite{bohm87, bohm85}. The capture processes calculated with these approaches show significant dependence on temperature, which is verified experimentally\cite{henry77} and is valid for shallow traps ($\sim k_\mathrm{B}T$) in equilibrium with the delocalized states. 

Deep traps ($\gg k_\mathrm{B}T$), on the other hand, had been treated in a variety of ways. Some works continue treating deep traps in the same manner as shallow ones\cite{Khalfin85}, while other works use a different approach. Lax\cite{lax60} noticed that inclusion of multiphonon transitions to the ground state in the modelling of deep traps yields cross-sections orders of magnitude too small. He showed\cite{Gummel57} that separation of carrier capture in two stages: ``binding'' of the electron to the trap, and the multi-phonon relaxation down the trap, leads to cross-sections of the correct order of magnitude. Similar notion was given by Curie for the reverse process of electron release (p.~145 in Ref.~\onlinecite{Richards63}), it was suggested that accumulating energy from absorbed phonons on the site is not enough to delocalize the trapped electron, and the probability to free an electron is described separately in the pre-exponential factor. Thus the frequency factor (and the release rate) should show no significant temperature dependence\cite{gibbs72}. 

The discrepancy and huge variation in approaches  to describe the underlying mechanisms of charge carrier (de)trapping (from semi-classics\cite{Markvart81} to variety of quantum mechanics models \cite{lax60,abakumov85,Das20}) led us to the presented work. 

In this manuscript we are mainly focused on the deeper traps ($\gg$k$_\mathrm{B}$T) and pre-exponential factor, thus we do not involve advanced forms of multi-phonon interactions. The release and capture of charge carriers are treated as a two-stage process. The first stage is the accumulation of phonons on the defect site. Here, the macroscopic thermodynamic limit is exploited for thermal release. The key role is given to the phonon occupation number which obeys Bose-Einstein statistics. The second stage is a resonant transition (tunneling) of an electron between the conduction band and a localized center (with no change in energy). The huge advantage of this approach is that it allows to consider release and capture as mutually complementary processes due to common absolute value of the hopping/transition matrix element. \textcolor{black}{Then the pre-exponential factor is also directly defined as a parameter reversed to capture coefficient, which will help us in clarifying its physical nature.} \textcolor{black}{Our goal is to provide a robust and simple model with the capability to connect the quantum picture with classical picture of frequency factors in thermoluminescence and of charge transfer phenomena.}

For ease of discussion around capture/release process we regularly use the example of Ce$^{3+/4+}$ in an oxide matrix - an ion with a single electron on an outer shell, and a popular choice for an efficient dopant in inorganic phosphors\cite{Nishiura11,Shah03}. The introduced model is validated by connecting it to the experimental data available in literature, and by derivation of all classic-model results discussed in the overview below as the limit case of the proposed model. Finally we discuss new insights into attractive/neutral Coulomb and ``compact'' traps and the temperature dependence of the capture process for deep traps.

\subsubsection*{Overview of the classical approach}\label{subsubsec:overview}

Regardless of their quantum nature, most charge transport processes may be described quantitatively in a semi-classical approach. This makes sense because kinetic equations normally regulate the behavior of a vast ensemble of particles, resulting in macroscopically visible consequences. At the same time the quantum nature of the processes is hidden in the  microscopic theoretical description of the constants entering the dynamic equations. 

Let us consider the capture and thermal release of an electron from the conduction band (CB) by a trap. The time-dependent concentration of available electrons in CB, $n_{e}$, follows a conventional rate equation, 

\begin{equation}
    \frac{dn_e}{dt}=A \cdot (N_{tr}-n_{tr}) \cdot n_e,
\label{chen1}
\end{equation}

\noindent where $A$ is a capture coefficient (in units of $\SI{}{\centi \meter^{3} \second^{-1}}$), $N_{tr}$ is the total concentration of capturing ions in the lattice, and  $n_{tr}$ is the number of occupied capturing centres, both in units of $\SI{}{\centi \meter^{-3}}$. 

Filled traps can also be thermally depleted. The thermal release of captured charge carriers to the CB is usually phenomenologically described with an Arrhenius expression \cite{arrhenius1889},

\begin{equation}
    \frac{dn_{tr}}{dt}= n_{tr} \cdot s \cdot e^{-\frac{E_a}{k_{\mathrm{B}}T}},
\label{chen2}    
\end{equation}

\noindent where $s$ is a frequency factor [s$^{-1}$], $E_a$ is the thermal activation barrier, [eV], $k_{\mathrm{B}}$ - Boltzmann constant, [eV~K$^{-1}$].

To govern the localization and de-localization processes an additional correlation between eq.~(\ref{chen1}) and (\ref{chen2}) is then added, introducing the detailed balance between trapping and de-trapping at sufficiently long times \cite{chen11, pierret87}. The straightforward maths results in the following expression:

\begin{equation}
    A=s\cdot N_c,
\label{chen3}    
\end{equation}

\noindent where $N_c$ is the ``effective density of the conduction band states'' \cite{chen11, pierret87}.

The capture coefficient $A$ is further evaluated in the framework of classical scattering theory. 
Considering an electron as a particle with small radius $r_{e}$ (typically of the order of the Thomson electron radius, $r_{e}=\alpha^{2}a_0$, with $\alpha$ being the electromagnetic fine structure constant and $a_0$ as the Bohr radius) migrating through a crystal, the capture coefficient is a product of the thermal velocity
and the trapping cross-section with an effective radius $a_0$ \cite{pierret87}:

\begin{equation}
    A=\upsilon_T \cdot \pi a_0^2
\label{chen4}
\end{equation}

The discussed approach leads to a well-known expression for the frequency factor $s$ as the capture rate per effective density of the conduction band states,

\begin{equation}
    s = \upsilon_T \frac{ \pi a_0^2}{N_c}.
\label{classfreqfac}
\end{equation}

In this classical description the frequency factor is interpreted as the number of times per second a captured electron interacts with the lattice phonons, multiplied by a transition probability\cite{mckeever88, bube78}. 
The maximum rate of the electron-phonon interaction is then related to typical phonon frequencies of 10$^{12}$-10$^{14}$~s$^{-1}$\cite{bos06}, while the transition probability and its microscopic origin generally lack a thorough description. 

The presented formulas form a basis for a reasonable interpretation and basic evaluation of the charge transport within a crystal, but they implicitly exclude its quantum nature\cite{stallinga11}. The semi-classical model \cite{chen21} also fails in a more advanced quantitative description \cite{frenkel38, nenashev18} or shows weaknesses for certain critical  questions. As such the eqs.~(\ref{chen4}),~(\ref{classfreqfac}) do not take into account the interaction between the charge carrier and the lattice. In addition, they cannot explain the experimentally observed huge variations in frequency factors between organic \cite{carr15} and inorganic\cite{Laguta12} materials, as well as the marked differences between ionic\cite{McKeever80, kawano20} and covalent \cite{Li18,vedda08,nikl07} compounds, p.~144 in Ref.~\onlinecite{Richards63}. A description of Coulomb-active (a trap aliovalent to the regular lattice ions) and Coulomb-neutral centers with capture coefficients differing by orders of magnitude\cite{chen11} also require an addition of the interaction potential to the model. Furthermore a discrepancy is observed for low-temperature traps (e.g. anti-sites\cite{Nikl05} in garnets): they have similar thermal depths, but frequency factors vary from $10^6$ to $\SI{d12}{\per \second}$ \cite{Drozdowski14, Brylew14}.

More specific aspects of the classical theory can be critically questioned in terms of an interpretation\cite{chen16} of the constants in eq.~(\ref{classfreqfac}). In particular, the interpretation of the frequency factor in a chemical kinetic framework is problematic. It does not readily allow to understand the experimentally observed variation of the frequency factor by several orders of magnitude for structural defects and impurities\cite{chen11, Drozdowski14, vedda08, nikl07, ueda15}. 

Another questionable aspect of the classical approach arises when one carries out time and temperature dependent experiments on gamma-ray-excited scintillators. In these experiments, the rise-time of the scintillation flash is reported to vary between 10~ps\cite{Moszynski98, Derenzo00, auffray16, gundacker18, Gundacker19, Gundacker20, Martinazzoli22} to 1-10~ns\cite{ Derenzo00, Seifert12, weele14, Zapadlik22}, and being independent of temperature in the range of 80-500~K\cite{weele14, weele15}. The scintillation rise time is the time that is required for the cascade of hot charge carriers to thermalize and localize on recombination sites. The loss of energy and thermalization of hot carriers occurs very rapidly ($<ps$), the rest of the time is spent by capture of thermalized carriers on the recombination sites \cite{rodnyi97}.
In the classical approach, the capture rate, eq.~(\ref{chen4}) is the product of the trapping cross-section, the trap concentration and thermal velocity. Temperature modifies the thermal velocity and with the equipartition theorem of thermodynamics a $\sqrt{T}$-dependence is anticipated \cite{mckeever88, chen69, bemski58, lax60, balarin79, gibbs72}. Accordingly, the capture rate is predicted to slow down to a value close to zero at cryogenic temperatures.  However, this $\sqrt{T}$-expectation is not well reproduced by experimental data \cite{weele15} indicating the failure of certain approximations.  

Following the discussion above, the main purpose of our work is to introduce a reliable analytical model that provides clear description of charge transport and capture mechanism, and is able to reproduce the main results of the semi- classical kinetic theory (in the spirit of Bohr’s correspondence principle).
\textcolor{black}{The model is useful for interpretation of experimental results around charge transfer and capture phenomena, e.g. thermoluminescence experiments. Frequency factors and capture cross-sections gain direct physical meaning and even allow to speculate about the spatial extent of the wavefunctions of the charged trap state. The model potentially allows to use thermoluminescence data as a validation tool for results obtained in \textit{ab~initio} calculations for localized states, in the same way as other experimental methods\cite{freysoldt2014}.}

\section{Model setup}\label{sec:modelsetup}

\subsection{Thermally assisted release}\label{subsec:thermassre}
We assume a spatial equilibrium configuration and consider thermal ionization of an electron from a localized level into the continuum of the CB, schematically depicted in Fig.~\ref{scheme}. Here we picture the 5d$_1$ (Ce$^{3+}$)$^\ast$-state and its thermal ionization as an example\cite{ueda19}. We split the process in two stages\cite{lax60, Richards63}: 1) electron-phonon interaction on the site (5d$_1$ Ce$^{3+}$-electron and the surrounding ligands) to gain the required activation energy, 2) resonant de-localization of an electron to the minimum of the CB, often being the $\Gamma$ point of the first Brillouin zone (BZ). The second stage does not involve any gain/loss of the electron energy and will be described by Fermi's golden Rule. The use of these two stages assumes that the system follows the Franck–Condon principle, thus the energy of phonons occupying the site consists of two parts with quasi instantaneous values: kinetic energy of ions in the lattice and potential energy of the electronic subsystem. According to the Born–Oppenheimer approximation, we can treat these two subsystems separately, where the ionic positions have to be accounted for as a quasi static multi-dimensional parameter, which is represented by the configurational coordinate $q$ on the abscissa in Fig.~\ref{scheme}. The effect of ion displacement on the electronic subsystem is depicted as a harmonic potential within the configurational coordinate diagram and shows that resonant tunneling is possible when the site is in its extreme position on the branch of the parabola (ions having vanishing kinetic energy).
\textcolor{black}{During tunneling the ions are frozen in their instantaneous (nearly-extreme) positions. After the charge transfer occurred, the ions relax to their new respective equilibrium positions due to change in local charge density, shown by blue (filled trap) and green (empty trap) parabolas in Fig.~\ref{scheme}.} This property is crucial for treating capture and release processes as mutually complementary.

\subsubsection{Stage\,I: Electron-phonon interaction}\label{subsubsec:stage1}

\begin{figure}[h!]
\centering
\includegraphics[scale=0.31]{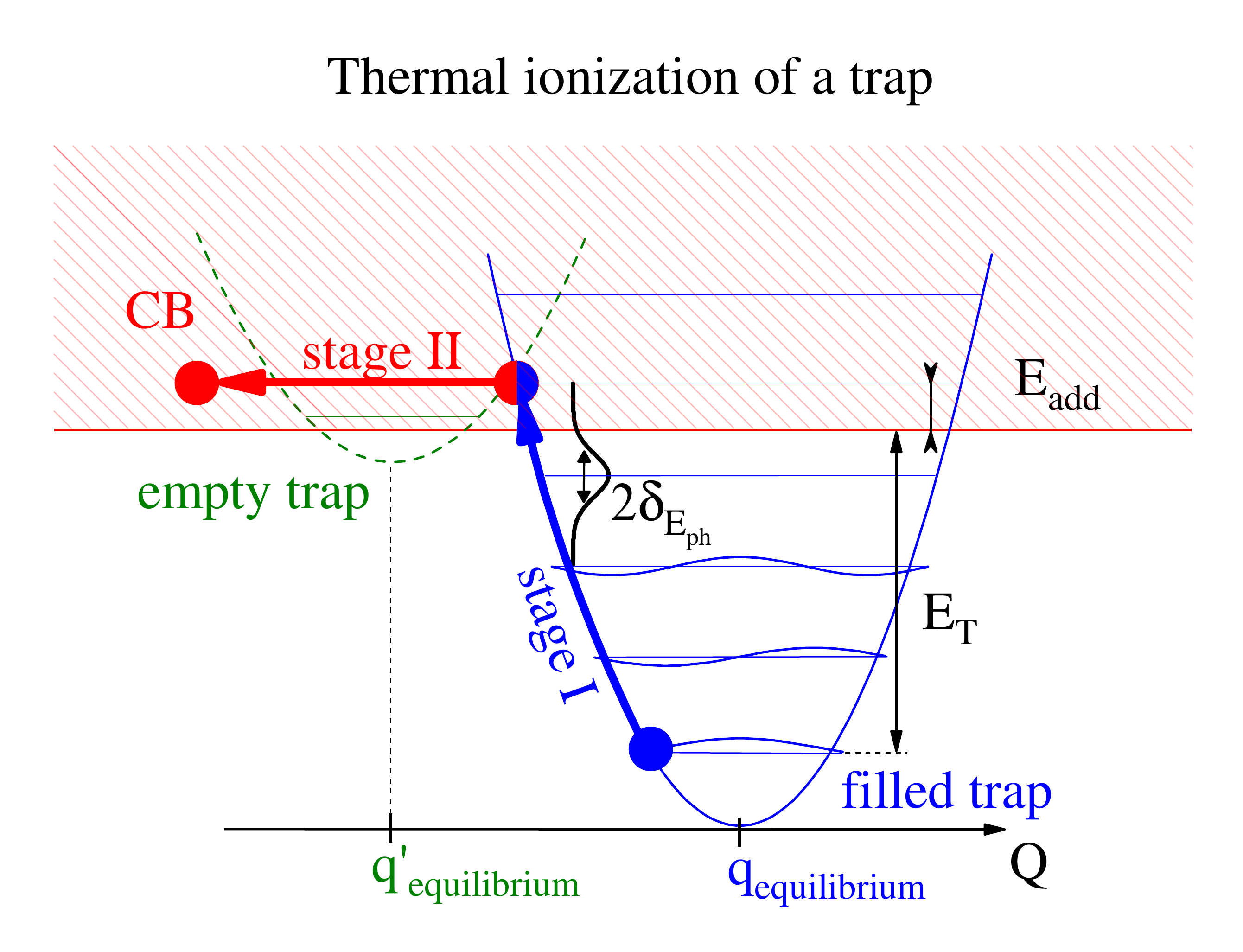}
\caption{The schematic diagram showing stages and states that form the thermally-assisted release of the electron from the trap site (or vice versa). \textcolor{black}{The blue parabola represents the filled trap,  the green one indicates the empty trap, $q_\mathrm{equilibrium}$ and $q'_\mathrm{equilibrium}$ are the equilibrium positions of the filled and empty defects, respectively.}  Stage~I is the thermal excitation of the electron (blue arrow). Stage~II is the resonant release (or resonant capture if the process is reversed) of the electron into the continuum of CB states (red arrow). Blue spheres show the localized electron, red sphere - the electron at the CB minimum (in reciprocal space). The energy deposited on the trap site by phonons is composed of two parts: thermal depth E$_{T}$ of the localized state and the additional energy E$_{\mathrm{add}}$ provided by potentially excessive amount of phonons. E$_{\mathrm{add}}$ in stage\,II  transforms to kinetic energy of the CB-electron. The broadening of the phonon levels is depicted as $2\delta_{E_{ph}}$. }
\label{scheme}
\end{figure}

For an escape into the CB, a localised electron must have an excess of energy that can be acquired thermally by interaction with phonons. Within this work, we will consider the presence of one effective optical phonon mode of energy $\hbar \omega_0$ for the given trapping centers. In the context of luminescent materials, this model often proves to be sufficiently valid, especially at temperatures far below the Debye temperature. The occupation number for the effective optical phonon mode \textcolor{black}{occupying an impurity} is given by the Planck factor\cite{tavgin90} (see Appendix~\ref{Appphonon}):

\begin{equation}
    \rho(n)=(1-\mathrm{e}^{-\frac{\hbar \omega_0}{k_{\mathrm{B}}T}}) \mathrm{e}^{-\frac{n \hbar \omega_0}{k_{\mathrm{B}}T}}
    \label{occnumdiscr}
\end{equation}

\noindent where $n$ is the occupation number of the specific optical phonon mode, $\hbar \omega_0$ is the energy of a specific optical phonon mode, and $T$ is the absolute temperature. 

The requirement on the minimum number of phonons that will align the site with the CB-bottom in terms of energy is $n_{\mathrm{min}}=\frac{E_T}{\hbar \omega_0}$ (see Fig.~\ref{scheme}). Since $n$ becomes very large, then the summation over phonon occupation numbers can be replaced by an integration over the continuous variable of energy :

\begin{equation}
    \frac{\mathrm{d}P (E_{\mathrm{ph}})}{\mathrm{d}E_{\mathrm{ph}}}=(1-e^{-\frac{\hbar \omega_0}{k_{\mathrm{B}}T}}) e^{-\frac{E_{\mathrm{ph}}}{k_{\mathrm{B}}T}} \frac{1}{\hbar \omega_0},
    \label{occnumcont}
\end{equation}

\noindent where \textcolor{black}{ $\frac{\mathrm{d}P (E_{\mathrm{ph}})}{\mathrm{d}E_{\mathrm{ph}}}$ is the differential probability to have onsite additional continuously defined energy of $E_{ph}$}, 
which replaces occupation number. $E_{ph}$ is the combination of the trap depth ($E_T$) and the extra energy from excess phonons present on the site ($E_{\mathrm{add}}$):  $E_{ph}=E_{\mathrm{add}} + E_T$, see Fig.~\ref{scheme}.

An additional feature is the finite width of the thermally excited state (see Appendix~\ref{Appsmearing}). The probability for the site to absorb $n$ effective $\hbar \omega_0$-phonons has to be regarded within a continuous width $E_\mathrm{D}\in[-\delta_{E_{ph}};\delta_{E_{ph}}]$. The largest effect \cite{Alivisatos88, johns16, Bok20} on the linewidth of a state in a solid is usually coming from inhomogeneous broadening due to disorder and morphological defects in a bulk material. The broadening is generally estimated to be about 0.1-0.2~eV \cite{johns16, Bok20}, this value also fits the Ce$^{3+}$-doped garnets example as its width of experimental absorption band at low temperature is $\approx \SI{0.2}{\electronvolt}$ \cite{bachmann09}. We assume that the contribution from phonons occupying the impurity ion ($E_{ph}$) and the broadening ($E_{D}$) are independent and thus, the differential thermal excitation probability \textcolor{black}{$\dd{P}(E_{\mathrm{ph}},E_\mathrm{D})$} is given by the two-dimensional function:

\begin{multline}
    \dd{P}(E_{\mathrm{ph}},E_\mathrm{D})=\qty((1-e^{-\frac{\hbar \omega_0}{k_{\mathrm{B}}T}}) e^{-\frac{E_{\mathrm{ph}}}{k_{\mathrm{B}}T}} \frac{1}{\hbar \omega_0} \dd{E}_{\mathrm{ph}}) \times \\ 
    \times \qty(\rho(E_\mathrm{D}) \dd{E}_\mathrm{D})
\label{phonassexcdos}
\end{multline}

\noindent The formula above characterizes the two-dimensional differential function for a localized center to have an energy of $E_{ph}+E_{D}$. Later, this expression is used as a probability density function to calculate thermodynamic average number of centers attempting to release electrons into the CB.

\subsubsection {Stage\,II: Resonant release of a localized e$^-$ to CB} \label{subsec:resrel}

\textcolor{black}{If enough phonons are collected on impurity site then the resonance conditions is achieved. This condition automatically satisfies the energy conservation law an we are able to } 
consider the second stage of the thermal ionization which is a resonant release of an electron from a thermally excited (Ce$^{3+}$)$^{\ast}$ ion to the CB minimum. The resonant release rate, $W_{if}$, is given by Fermi's golden rule,

\begin{equation}
    W_{if}=\frac{2 \pi}{ \hbar} |\bra{f}\hat{V}\ket{i}|^2 \rho(E_f) = \frac{2 \pi}{ \hbar} \abs{\mathcal{M_{\mathrm{real}}}}^2 \rho(E_f),
\label{fermigoldenrule}
\end{equation}

\noindent where $\hat{V}$ is the static perturbation created by phonons, $\ket{i}$ and $\ket{f}$ - are single particle wavefunctions of the initial and final states and  $\rho(E_f)$ is the density of final states. $\abs{\mathcal{M_\mathrm{real}}}$ denotes the electronic interaction matrix element. The wavefunctions below are normalized to the volume of the supercell $V_{\mathrm{sc}}$ (due to presence of dopants the enlarged unit cell volume equals to inverse impurity concentration $V_{\mathrm{sc}}=N^{-1}_{\mathrm{tr}}$), thus we rewrite the interaction matrix element as follows:

\begin{equation}
    \abs{\mathcal{M}_{\mathrm{real}}}^{2}= \frac{V_{\mathrm{sc}}}{V_\mathrm{s}} \abs{\mathcal{M}_{\mathrm{model}}}^{2} 
\end{equation}

\noindent where $V_\mathrm{s}$ is the volume of a sample, $\mathcal{M}_{\mathrm{\mathrm{model}}}$ is the Coulomb matrix element for wavefunctions of localized and CB states.

The resonant release rate depends on the density of final states, which are reached by the electron during the phonon-assisted excitation. In particular, the extra energy above the thermal depth of a localized level is of interest and describes the kinetic energy of the released electron acquired by annihilation of phonons,

\begin{equation}
    E_{CB}^{kin}=(E_{ph}+E_{D})-E_T
\end{equation}

\noindent with the combined phonon energy $E_{ph}$, level broadening $E_{D}$ and thermal depth of the localized level $E_T$ have been introduced before in eq.~(\ref{phonassexcdos}). Thus, the rate of resonant release $W_{\mathrm{res\,rel}}$ depends on this additional energy $E_{CB}^{kin}$ and can be written as follows:

\begin{equation}
    W_{\mathrm{res\,rel}}(E_{CB}^{kin})=\frac{2 \pi}{ \hbar} \frac{V_{\mathrm{sc}}}{V_\mathrm{s}} \abs{\mathcal{M}_{\mathrm{\mathrm{model}}}}^{2}   \rho(E_{CB}^{kin}),    
\end{equation}
\noindent where $\rho(E_{CB}^{kin})$ is the part of CB density of states that is accessible to the electron with additional energy $E_{CB}^{kin}$,  

\begin{multline}
    W_{\mathrm{res\,rel}}(E_{CB}^{kin})=\frac{2 \pi}{ \hbar} \frac{V_{\mathrm{sc}}}{V_\mathrm{s}} \abs{\mathcal{M}_{\mathrm{\mathrm{model}}}}^{2}  \times \\ \times \frac{2V_\mathrm{s}}{(2\pi)^3} \int_{\Omega}\dd[3]{k} \delta(E-E_{CB}^{kin}),
\label{wresrelint}
\end{multline}

\noindent and $\delta(E-E_{CB}^{kin})$ is the Dirac delta functional ensuring the resonance condition.

Replacement of $k$ and $dk$ with $E$ and $dE$ in eq.~(\ref{wresrelint}), respectively, allows to perform the integration over the Dirac delta functional. Under the assumption that de-localized electrons have nearly-free parabolic dispersion with an effective mass $m^\ast$, the final expression for resonant release rate is then written as:

\begin{multline}
    W_{\mathrm{res\,rel}}(E_{CB}^{kin})=\frac{2\pi}{\hbar} V_{\mathrm{sc}} \abs{\mathcal{M}_{\mathrm{\mathrm{model}}}}^{2} \times \\ \times \frac{8\pi}{2} \qty(   \frac{2 m^\ast}{h^2})^\frac{3}{2} \sqrt{E_{CB}^{kin}}
\label{wresrel}
\end{multline}

\noindent It is important that eq.~(\ref{wresrel})  specifies the rate of resonant release only for electrons having the excess energy in the vicinity of the value "$E_{CB}^{kin}$". To generalize the expression we need to average it over the ensemble of ionizing centers having different values of $E_{CB}^{kin}$.

\subsubsection{Ionization rate derivation}

Integration of the resonant release rate ($W_{\mathrm{res\,rel}}$) over the phonon occupation distribution ($P(E_{\mathrm{ph}},E_\mathrm{D})$) (obtained in section~\ref{subsubsec:stage1}, eq.~(\ref{phonassexcdos})) gives the ionization rate ($\gamma_{\mathrm{ion}}$) for the impurity site:

\begin{multline}
    \gamma_{\mathrm{ion}} = \int  W_{\mathrm{res\,rel}} dP =  \frac{2 \pi}{ \hbar} V_{\mathrm{sc}} \abs{\mathcal{M}_{\mathrm{\mathrm{model}}}}^{2} \times
    \\
    \times \frac{8\pi}{2} \qty(\frac{2 m^\ast}{h^2})^\frac{3}{2}  \int\limits_{E_{CB}^{kin}\ge 0 } \sqrt{E_{CB}^{kin}} \times dP(E_{\mathrm{ph}},E_\mathrm{D})
    \label{resrelgenview}
\end{multline}

We perform the integration in eq.~(\ref{resrelgenview}) numerically (see Appendix~\ref{AppInt} for details) and obtain the following:

\begin{multline}
    \gamma_{\mathrm{ion}} =  \frac{2\pi}{\hbar} V_{\mathrm{sc}} \abs{\mathcal{M}_{\mathrm{\mathrm{model}}}}^{2} \frac{1}{\delta_{E_{ph}}} \times \\  \times K_\mathrm{num} N_c e^{-\frac{E_T-\frac{\delta_{E_{ph}}}{2}}{k_\mathrm{B}T}}
\label{resrelfinres}
\end{multline}

\noindent where $K_\mathrm{num} \approx $1 for almost any physically relevant pair of phonon energy $\hbar \omega_0$ and uncertainty broadening $\delta_{E_{ph}}$. $N_\mathrm{c}$ is the effective density of states available for electrons in the CB,

\begin{equation}
    N_\mathrm{c} = \frac{ (2 \pi m^\ast k T) ^\frac{3}{2}  }  {h^3} = \lambda^{-3}_{\mathrm{dB}},
    \label{eq:nc}
\end{equation}

\noindent with $m^\ast$ as the effective electron mass in the CB and $h$ as  Planck's constant. $\lambda_{\mathrm{dB}}$ denotes the thermal de Broglie wavelength of the electrons in the CB.

We can then incorporate the whole combination of coefficients and variables in front of the exponent in eq.~(\ref{resrelfinres}) into a pre-exponential factor $s$ (units of s$^{-1}$):

\begin{eqnarray}
    \gamma_{\mathrm{ion}} =  s \cdot e^{-\frac{E_T-\frac{\delta_{E_{ph}}}{2}}{k_\mathrm{B}T}}
\label{shortrate}
\end{eqnarray}

\begin{eqnarray}
    s =  \frac{2\pi}{\hbar} V_{\mathrm{sc}} \abs{\mathcal{M}_{\mathrm{\mathrm{model}}}}^{2} \frac{1}{\delta_{E_{ph}}} K_\mathrm{num} N_c 
\label{freqfactor}
\end{eqnarray}

It is noteworthy that compared to the classic formula, the Boltzmann factor in eq.~(\ref{shortrate}) contains an effective activation barrier in a slightly more complex form: $E_a=E_T-\frac{\delta_{E_{ph}}}{2}$. According to our model, the thermal activation barrier ($E_a$) is actually slightly smaller than the plotted energy gap between the bottom of CB and the trap level ($E_T$, Fig.~\ref{scheme}) and reduced by the uncertainty term of $\frac{\delta_{E_{ph}}}{2}$. This term appears because a localized level with the depth less than a half of its broadening cannot trap carriers at any temperature.

Eq.~(\ref{freqfactor}) depicts the parametrization of the pre-exponential factor as the release rate and in the next section we will highlight the physical interpretation of $s$ [s$^{-1}$].

\subsection{Resonant capture}

Here we consider the resonant capture of electron from the CB at a Coulomb-active impurity site (e.g. at Ce$^{4+}$ in YAG:Ce). This is the reverse process to stage~II shown in Fig.~\ref{scheme} and described in subsection~\ref{subsec:resrel}. 

The ions charged w.r.t the lattice create a static perturbing potential, which is the same as the potential for resonant release. The capture rate is again assessed by Fermi's golden rule:

\begin{equation}
     W_{\mathrm{res\,capt}}= \frac{2 \pi}{ \hbar} \frac{V_{\mathrm{sc}}}{V_\mathrm{s}} \abs{\mathcal{M}_{\mathrm{model}}}^{2} \rho(E_\mathrm{f}),
\end{equation}

The vacant localized shell of the trap (Ce$^{4+}$ in our example case) available for capture is affected by the same uncertainty broadening of the state as during resonant release. 
Thus, the density of final states in Fermi's golden rule can be approximated using the same broadening contribution $E_\mathrm{D}\in[-\delta_{E_{ph}};\delta_{E_{ph}}]$ from the model probability density function (see Fig.~\ref{scheme} and Appendix~\ref{Appsmearing}):

\begin{equation}
     \rho_{\mathrm{res\,capt}}(E)=\frac{3}{4} \frac{ N_{\mathrm{tr}}V_\mathrm{s}}{\delta_{E_{ph}}}
\end{equation}

\noindent where $N_{\mathrm{tr}}V_\mathrm{s}$ is the total number of traps in the sample available for electron capture. 
Thus, the final formula for resonant capture is:

\begin{equation}
    W_{\mathrm{res\,capt}}=\frac{2 \pi}{ \hbar} V_{\mathrm{sc}} \abs{\mathcal{M}_{\mathrm{model}}}^{2} \frac{3}{4}\frac{N_{\mathrm{tr}}}{\delta_{E_{ph}}},
    \label{finresrate}
\end{equation}

\noindent which we can re-write in the form of the capture coefficient
$A_{\mathrm{res}}$ that does not depend on Coloumb-attractive trap concentration:

\begin{equation}
    A_{\mathrm{res}}=W_{\mathrm{res\,capt}}/N_{\mathrm{tr}}=\frac{2 \pi}{ \hbar} V_{\mathrm{sc}} \abs{\mathcal{M}_{\mathrm{\mathrm{model}}}}^{2} \frac{3}{4} \frac{1}{\delta_{E_{ph}}}.
\label{finrescapt}
\end{equation}

It is noteworthy that the capture coefficient $A_{\mathrm{res}}$ only depends on the interaction matrix element, broadening $\delta_{E_{ph}}$ of the localized level and the supercell volume $V_{\mathrm{sc}}$. Accordingly, from eq.~(\ref{freqfactor}), the pre-exponential factor $s$ follows as

\begin{equation}
    s = A_{\mathrm{res}}  N_{\mathrm{c}} \qty{\frac{4}{3} K_\mathrm{num}}
\label{finnumfreq}    
\end{equation}

\noindent with the dimensionless coefficient $K_\mathrm{num} \approx 1$ as a result of the numerical integration procedure and inclusion of the probability density function. Thus, we arrive at the general scaling 

\begin{equation}
    s \approx A_{\mathrm{res}}  N_{\mathrm{c}}, 
\label{finfreq}    
\end{equation}

\noindent in agreement to the classic approach (see eq.~(\ref{chen3})). However, our derivation offers a microscopic interpretation for the frequency factor.

The details of our derivations can be found in Appendices.

\section{Results}

The classical interpretation of an electron capture starts with the assumption that the capture radius of a neutral trapping center equals its geometric radius\cite{chen11}. Orders of magnitude variations of the cross-section for centers having a non-zero net charge w.r.t the lattice are justified by the presence of attractive or repulsive potentials. However, the effect of these potentials is scarcely treated in detail. To cover this gap we consider the typical ranges of microscopic parameters governing the thermal release/capture within our model and offer a physical interpretation. We start with calculations of the capture coefficient $A_{\mathrm{res}}$.

\subsection{Coulomb attractive center}\label{sebsec:coulomb}

\begin{figure}[h!]
\centering
\includegraphics[scale=0.31]{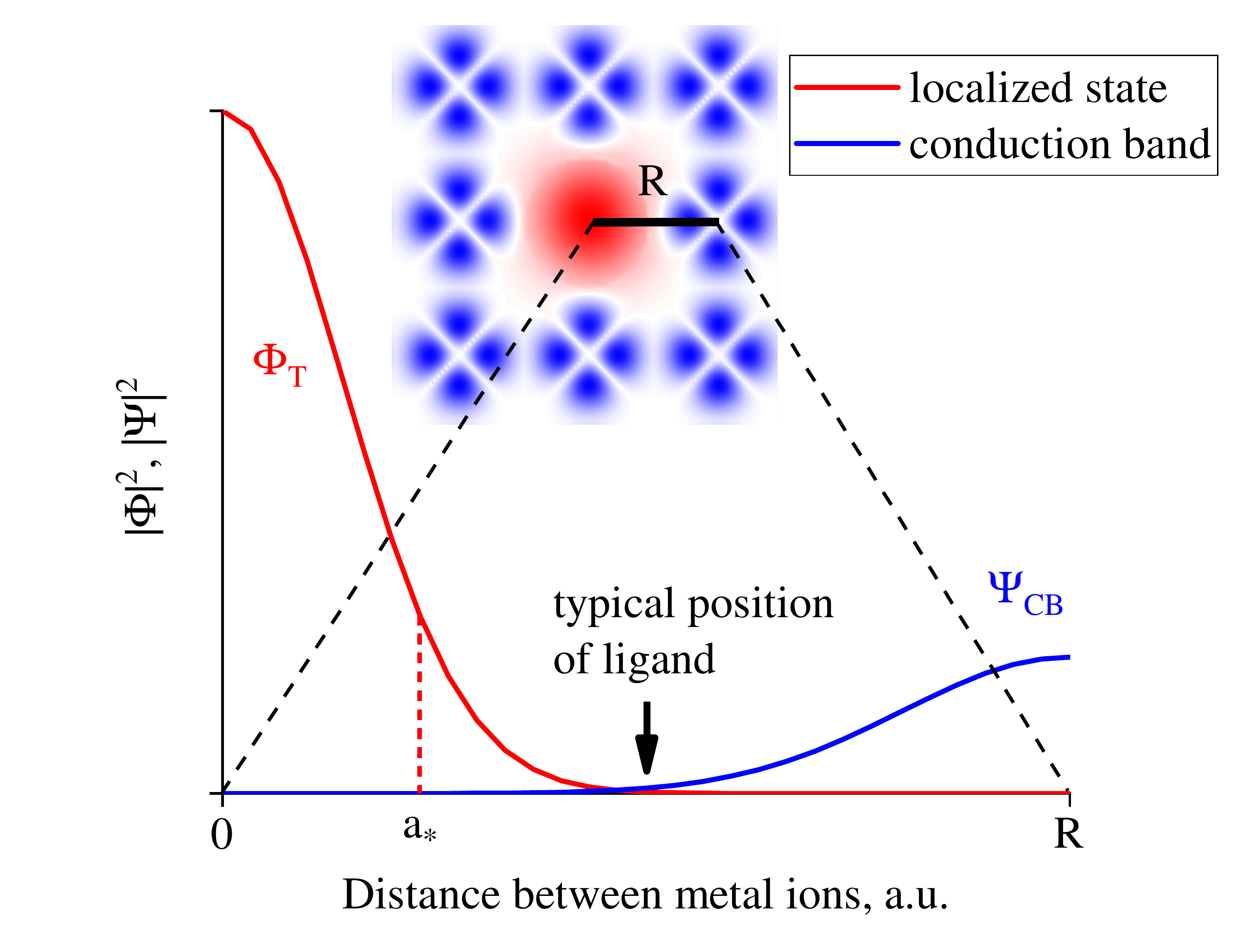}
\caption{Sketch of an overlap between localized and conduction band states presented as squares of wave functions' absolute values. Localized state (red curve) is formed by hybridized orbitals of impurity atom of metal and ligands (not shown for simplicity). The effective radius of the localized state is denoted as $a_{\ast}$. Conduction band states (blue curve) occupy orbitals of the regular lattice ions. An impurity ion (red) in the pierced lattice of regular ions (blue) together form the supercell, shown in the inset. R is the characteristic distance between the trap and regular ions nuclei.}
\label{sketchovlp}
\end{figure}

Let us consider the spatial representation of the wavefunctions forming a Coulomb attractive defect in a regular lattice (e.g. Ce$^{4+}$ in a YAG).
For the delocalized CB states we assume homogeneously distributed plane waves with $k \rightarrow 0$ pierced in the vicinity of an impurity. The impurity, for simplicity, is characterised by a Gaussian orbital with characteristic radius of $a_\ast$. The distance between pierced plane-wave cutoff and impurity nucleus is about $R$ (typical inter-metal distance in oxide insulators). We assume that inside the sphere of radius $R$ the CB state decays equivalently to the Gaussian orbital, see Fig.~\ref{sketchovlp}.

Consequently, the wave function $\Phi_{T}$ of the capturing center can be written as

\begin{equation}
    \Phi_{T} (r) = \qty{\frac{2}{\pi a_\ast^2}}^{3/4} e^{-(\frac{r}{a_\ast})^2}
\label{impurapprox}
\end{equation}

\noindent and the wave function for the CB states $\Psi_{CB}$ inside the radius $R$ is

\begin{equation}
    \Psi_{CB} (r) \approx \frac{1}{\sqrt{V_{\mathrm{sc}}-4/3\pi{R}^3}} e^{-(\frac{r-R}{a_\ast})^2}, 
\label{CBapprox}
\end{equation}

Substitution of the wave functions and Coulomb potential (neglecting screening effects) in eq.~(\ref{finrescapt}) for the capture coefficient gives

\begin{multline}
    A_{\mathrm{res}}=\frac{2 \pi}{ \hbar} \frac{V_{\mathrm{sc}}}{\delta_{E_{ph}}}  \Bigl|  \int \limits_{0}^{R}     \qty{\frac{2}{\pi a_\ast^2}}^{3/4} e^{-(\frac{r}{a_\ast})^2} \frac{kq^2}{r} \times  \\   \times  \frac{1}{\sqrt{V_{\mathrm{sc}}-4/3\pi{R}^3}} e^{-(\frac{r-R}{a_\ast})^2}    4 \pi r^2 \dd[]{r}  \Bigl| ^2
\label{estrate}
\end{multline}

\noindent \textcolor{black}{where $q$ is the electron charge}.

The analytical solution of expression (\ref{estrate}) with $\text{Erf}[\frac{R}{\sqrt{2}a_\ast}] \approx 1$ gives:

\begin{eqnarray}
    A_{\mathrm{res}}=\frac{2 \pi}{ \hbar} \frac{V_{\mathrm{sc}}}{\delta_{E_{ph}}} \frac{16\sqrt{2} \pi^{3/2} \kappa^2 e^{-\kappa^2} (Ry^\ast)^2 a_\ast^3 }{ V_{\mathrm{sc}}-4/3\pi{R}^3},
\label{anestrate}
\end{eqnarray}

\noindent where $\kappa=R/a_\ast$ is the ratio of the inter-metal distance ($R$) and trapping orbital effective radius ($a_\ast$), $Ry^\ast$ is the Rydberg-like energy where Bohr radius is substituted by $a_\ast$.  Assuming that the volume of the supercell is much larger than the volume occupied by the trap ($V_{\mathrm{sc}} \gg 4/3 \pi R^3$), and reordering dimensional and dimensionless co-factors one obtains:

\begin{equation}
    A_{\mathrm{res}}=  16\sqrt{2} \pi^{\frac{3}{2}} \kappa^2 e^{-\kappa^2}\times
    \frac{1}{\frac{\delta_{E_{ph}}}{Ry^\ast}} \times 
    \frac{Ry^\ast}{\hbar} a_\ast  \times  
    \pi a_\ast^2
\label{finestrate}
\end{equation}

We can divide eq.~(\ref{finestrate}) into four distinct parts: 
\begin{itemize}
    \item a dimensionless factor ($16\sqrt{2} \pi^{\frac{3}{2}} \kappa^2 e^{-\kappa^2}$) characterizing overlap between the trap and the CB.  The $\kappa$ is the ratio between inter-metal distance and effective decay constant of hybridized atomic orbital;
    \item a dimensionless factor ($\frac{1}{\frac{\delta_{E_{ph}}}{Ry^\ast}}$) characterising uncertainty broadening of the vacant trap level;
    \item a ``velocity'' - like term $\frac{Ry^\ast}{\hbar} a_\ast$ with dimensions of [cm$\cdot$s$^{-1}$];
    \item a geometric cross-section of the capturing orbital $\pi a_\ast^2$ [cm$^{2}$].
\end{itemize}

Upon insertion of estimates for these factors in eqs.~(\ref{impurapprox})-(\ref{finestrate}) show that despite the presence of the effective cross-section and velocity-like factor, the capture coefficient and frequency factor significantly depend on quantum microscopic parameters of the trapping site, namely the interaction potential, local overlap of the CB with localized states in the vicinity of the trap, and energy broadening of the trap level.

For the example of Ce$^{4+}$ in YAG:Ce, we implement some typical values for uncertainty broadening of $\delta_{E_{ph}}$=0.1~eV\cite{bachmann09}, and inter-metal distance in garnet lattice of $R$=3.7~\AA\cite{garcia11}.
\textcolor{black}{For the effective trap radius $a_\ast$ we take an average between Y$^{3+}$ (1.02~\AA) and Ce$^{3+}$ (1.14~\AA) ionic radii in (8)-fold coordination\cite{Shannon76}, as such $a_\ast = 1.08\text{~\AA}$. Putting these numbers into eq.~(\ref{finestrate}) we obtain an estimate (with an order of magnitude precision) for the capture coefficient for Ce$^{3+}$ 5d orbital acting as a trap:}

\begin{multline}
    A_{\mathrm{res}} \approx  10^{-2} \times
    10^2 \times
    10^8  \mathrm{cm\cdot s^{-1}} \times  
    3\cdot 10^{-16} \mathrm{cm}^2 = \\ = 3\cdot 10^{-8} \mathrm{cm^3s^{-1}}
\label{numbersestrate}
\end{multline}

Thus with a simple model using Gaussian atomic orbitals and the Coulomb potential, we arrive at the same order of magnitude for the capture coefficient and matrix element as determined from experiments\cite{khanin19}.
Considerable variations are expected in the overlap between localized states and continuum depending on the interface and decay behaviour of the respective wave functions. More accurate modelling requires well-defined trapping sites and lattice properties. 

\subsection{Model for neutral trap}

\begin{figure}[h!]
\centering
\includegraphics[scale=0.31]{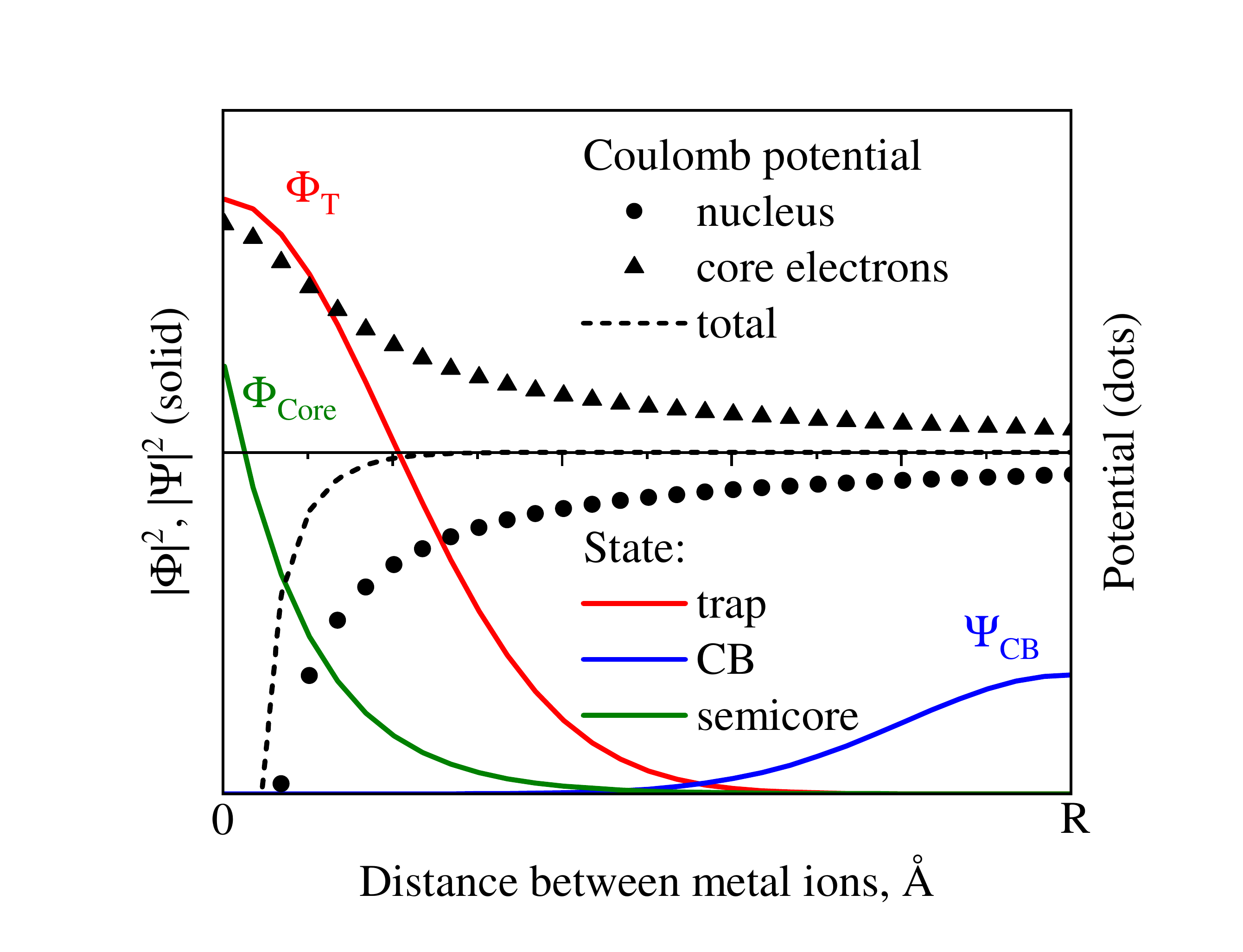}
\caption{Schematic illustration of  the electron charge densities (solid curves) for localized ($\Phi_{\mathrm{T}}$), conduction band ($\Psi_{\mathrm{CB}}$) and semicore ($\Phi_{\mathrm{Core}}$) states. Coulomb attractive potential (black dots) of the impurity ion nucleus and Coulomb repulsive potential (triangles) of the impurity core shell compensate each other, creating compact effective single particle potential (dashed curve).}
\label{neutral}
\end{figure}

The model presented above to assess the matrix element for the case of Coulomb attractive capturing center (e.g. Ce$^{4+}$ in garnets) may be naively extended to the case of a neutral trap. A doping ion, which is isovalent to the substituted ion site of the host compound usually has localized semicore electrons that shield the Coulomb behaviour of the single particle potential. For that, we need to introduce additional parameter $a_{\mathrm{core}}$ characterizing spatial distribution of the localized semicore level wavefunction in the following form

\begin{equation}
    \Phi_{\mathrm{core}}(r)=\frac{1}{\sqrt{\pi a_{\mathrm{core}}^3}}e^{-\frac{r}{a_{\mathrm{core}}}}
\label{semicorewf}
\end{equation}

\textcolor{black}{Below we get the modified operator for resonant capture as a sum of two electrostatic potentials: the first ($V_{\mathrm{Coulomb}}$) goes for nuclear point charges, the second ($V_{\mathrm{core}}$) goes for electronic charges distributed over the core orbital:}

\begin{multline}
    V_{\mathrm{neutral}}(r)= V_{\mathrm{Coulomb}}+V_{\mathrm{core}}=-\frac{kq^2}{r} + \\ + \frac{kq^2  
    e^{-\frac{2r}{a_{\mathrm{core}}}}}{r}\left( e^{\frac{2r}{a_{\mathrm{core}}}}- \frac{r}{a_{\mathrm{core}}}-1 \right)
\label{semicorepot}
\end{multline}

The sketch of electron charge density and related potentials is depicted in Fig.~\ref{neutral}. The new effective potential acts on a much shorter range than the initial Coulomb potential due to the distributed electronic density of the inner semicore state. Estimates show that the matrix elements ratio for Coulomb attractive and neutral centres with the semicore characteristic radius of $a_{\mathrm{core}}\approx$0.75~\AA\ is about:

\begin{eqnarray}
    \frac{A_{\mathrm{Coulomb}}}{A_{\mathrm{neutral}}} \propto \left( \frac{\mathcal{M}_{\mathrm{Coulomb}}}{\mathcal{M}_{\mathrm{neutral}} } \right)^2 \approx 10^{3}  
\label{neutraleff}
\end{eqnarray}

The tiny variations of the semicore shell radius leads to the orders of magnitude variations of the matrix element value.

\subsection{Model for a compact trap}

Let us consider a compact point defect with reduced distributed electron density (e.g. an impurity of small ionic radius, an anti-site defect or a vacancy). Intuitively, the interaction potential for trapping/detrapping on such a site has to be significantly reduced.

The reduction in transition matrix element is reflected in the resonant capture coefficient, eq.~(\ref{finestrate}), mainly in the term $e^{-\kappa^2}$ describing the overlap between the wavefunctions of the defect and the CB. Here, $\kappa$ is the ratio between the metal-metal distance $R$ and the effective radius of impurity $a_\ast$. For the sake of demonstration, let us assume a decrease of $20\%$ of the effective radius $a_\ast$. This changes the ratio of the capture coefficients for the normal $\qty(A^{n})$ and small $\qty(A^{s})$ traps to:

\begin{eqnarray}
    \frac{A^n}{A^s} \propto  
    e^{-R^2\left(\left(\frac{1}{a_\ast^n}\right)^2 - \left(\frac{1}{a_\ast^s}\right)^2\right)} \approx 10^3
\label{smalltonormimp}
\end{eqnarray}

\noindent This yields the intuitive result that a point defect that is poorly coupled to its surrounding has a reduced capture cross-section and frequency factor by several orders of magnitude. This interpretation directly explains the huge variations in the frequency factor observed for different types of traps in a given compound \cite{Drozdowski14, vedda08, nikl07}.

\subsection{Thermal dependence of the capture coefficient and frequency factor}
    
Besides their physical meaning the most promising feature of the explicit formula in eq.~(\ref{finrescapt}) for the capture coefficient and frequency factor is an opportunity to assess their temperature behaviour. Let us employ a quite simple approach and use the analytically derived formula from the model of overlapping localized and conduction states (see eq.~(\ref{finestrate})). The key element of the estimation is the dimensionless parameter $\kappa$ characterising the ratio between inter-ion distance and capturing orbital radius. Both are obviously affected by the distortion caused by the propagation of a phonon.

\begin{figure}[h!]
\centering
\includegraphics[scale=0.31]{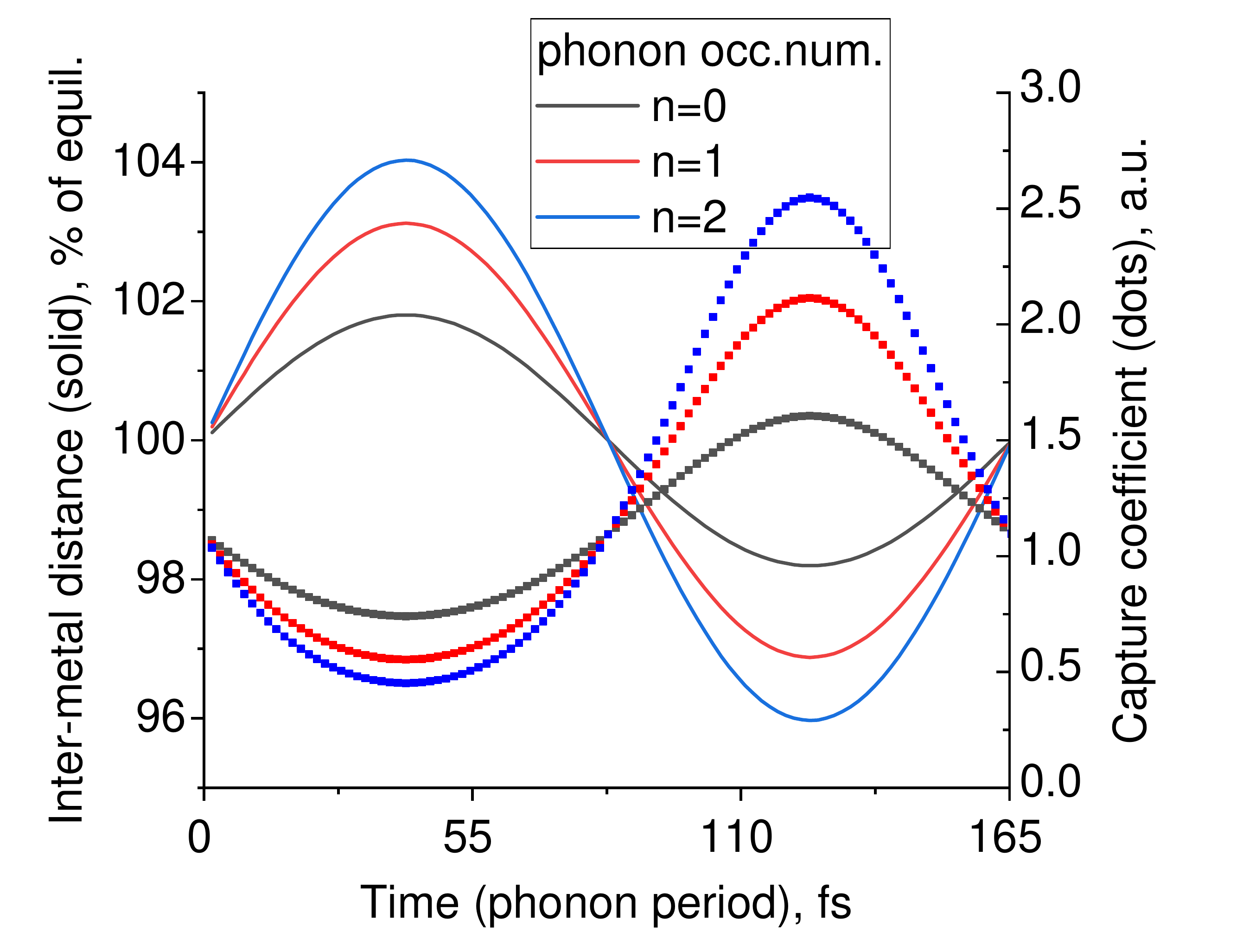}
\caption{The variation of the inter-ionic distance in a garnet structure (solid) and the related modelled capture coefficient (dots). Both are estimated during the phonon period when a different number of phonons (0, 1, 2 - black, red, blue) occupies the potential of the trapping site.}
\label{dynamics}
\end{figure}

So far we have used phonons only as a source of energy allowing the trapping site to achieve resonance with the conduction band states (in section \ref{subsubsec:stage1}). However, the matrix element in eq.~(\ref{estrate}),~(\ref{anestrate}) significantly depends on the distance between ions that form interacting states. Thus, one can observe additional, more delicate temperature-related effects. In a regular crystal structure the inter-ionic distance is actually time-dependent based on the thermal excitation of lattice vibrations, and can be approximated as a harmonic oscillator:

\begin{equation} 
     R(T,t)=R_0+ A_{max} \sin(t)
\label{instdist}
\end{equation}
\noindent where $R_0$ is the equilibrium distance between the ions at $T = \SI{0}{\kelvin}$,  $A_{max}$ is an amplitude of phonon-assisted fluctuations of ionic positions, where the last term is a function of temperature and phonon occupation number.

The amplitude of the ion oscillations can be quantized in the following manner:

\begin{equation} 
     \frac{kA_{max}^2}{2}=\hbar \omega_0(n+\frac{1}{2}) \, \rightarrow\, A_{max}=\sqrt{\frac{2 \hbar \omega_0(n+\frac{1}{2})}{ k}},
\end{equation}

\noindent where $n$ is the phonon occupation number and $k$ is the bond stiffness. The latter obeys mechanistic approach:

\begin{equation} 
     k=M_{ion}\omega_0^2
\end{equation}

\noindent where $M_{ion}$ is the mass of an ion. Now we are able to rewrite the time-dependent variation of the inter-ionic distance as

\begin{equation} 
     R(T,t,n)=R_0+ \sqrt{\frac{2 \hbar (n+\frac{1}{2})}{M_{ion} \omega_0}} \sin(t)
\end{equation}

We derive the time-dependent inter-ionic distance and the \textcolor{black}{effective capture coefficient}, taking into account the variation of the wavefunction overlap as represented by the dimensionless "$\kappa$" parameter according to eq.~(\ref{finestrate}),

\begin{equation}
A_{\mathrm{res}} \propto \kappa^2 e^{-\kappa^2} = (R/a_\ast)^2e^{-(R/a_\ast)^2}, 
\end{equation}

For the sake of clarity we plot the behaviour of this overlap-related part of the capture coefficient for the specific example of YAG. In this lattice, the Y$^{3+}$ ions have a mass of 39~a.u., while we estimate the frequency of  \textcolor{black}{an effective phonon mode} mode by $\omega_0\approx$ 3$\cdot$10$^{13}$~s$^{-1}$ \cite{lin18,bachmann09}. We adopt the effective radius of overlapping orbitals and inter-ionic distance mentioned in eqs.~(\ref{finestrate}) and (\ref{numbersestrate}) of $a_{\ast}$=1.08~\AA\, and R=3.7\AA. Insertion of these values into eq.~(\ref{instdist}) shows that the oscillatory changes of the inter-ionic distance can contribute to a few percent of $R_{0}$, (see solid lines in Fig.~\ref{dynamics}). These tiny variations in the ion positions lead to significant changes (factor of $\times$5) in the capture coefficient (Fig.~\ref{dynamics}, dotted lines). Thus, we deduce that temperature affects the trap capture coefficient by an increased oscillatory motion of the ions surrounding the trapping site.

\begin{figure}[h!]
\centering
\includegraphics[scale=0.31]{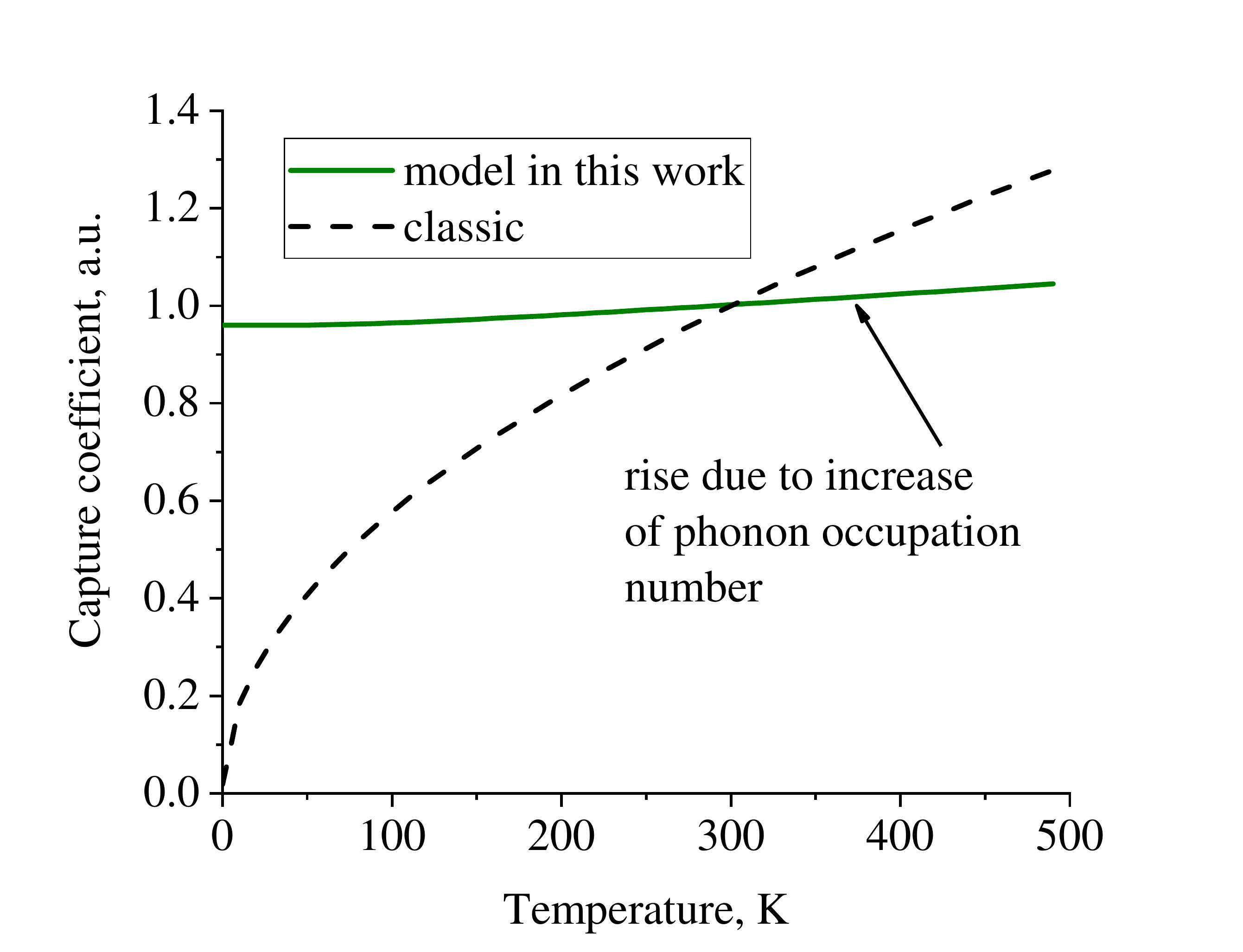}
\caption{The thermal dependence of the capture coefficient for the proposed and classical models. The proposed result is obtained by averaging of the time dependent capture coefficient (see eq.~(\ref{arescaptaveraging})) over time and occupation numbers, while the classical result scales with $\sqrt{T}$ \cite{chen11}. Both curves are normalized to the unity at room temperature. The main difference is that the quantum consideration predicts the natural non-zero limit of capture coefficient for low temperatures, while classic model suggests vanishing of a capture process due to moderation of carriers velocity.}
\label{tempdepbasic}
\end{figure}

We average the capture coefficient over the phonon period and phonon occupation number and obtain its temperature dependence in the thermodynamic limit as:

\begin{equation}
   \overline{A}_{res}(T) \propto \sum_{n=0}^\infty \rho(n) \frac{1}{\theta_{\hbar\omega_0}} \int_0^{\theta_{\hbar\omega_0}} A_{\mathrm{res}}(T,t,n)\dd[]{t}
\label{arescaptaveraging}
\end{equation}

\noindent where $\rho(n)$ is the phonon occupation number according to eq.~(\ref{occnumdiscr}), $\theta_{\hbar\omega_0}$ is the period of ionic oscillations, and $\overline{A}_{res}(T)$ is the averaged capture coefficient normalized to unity at room temperature. The expected temperature dependence of $\overline{A}_{res}(T)$ according to our model is depicted as green solid line in Fig.~\ref{tempdepbasic}. At very low temperatures the quantum picture predicts a constant non-zero value for the capture coefficient, that slowly rises with elevating temperature due to increased phonon occupation. Constant capture cross-section at cryogenic temperatures had been observed experimentally\cite{mitonneau79}.

A significantly different dependence is derived from the semi-classical approach. For that, the temperature behavior of the (de)trapping process is connected to thermal velocity. The rate of CB electron localization is assumed to follow a $T^{p}$ dependence\cite{bonch68}, or, in particular, the $\sqrt{T}$ temperature dependence \cite{chen11}:

\begin{equation}
   \overline{A}_{\mathrm{classic}}(T) \propto 
   \sqrt{\frac{8k_{\mathrm{B}}T}{\pi m^\ast}}
   \cdot \pi a_\ast^2
   \label{aclassicT}
\end{equation}

\noindent where $m^\ast$ is the effective mass of the electron, and $a_\ast$ is the trap effective radius. The $\sqrt{T}$ dependence explicitly suggests that the capture rate significantly slows down at low temperatures until no capture would occur at around cryogenic temperatures, see dashed curve in Fig.~\ref{tempdepbasic}.

The semi-classical and quantum approaches show discrepancies in the variation of the capture coefficient: the latter predicts significantly smaller changes, which are well supported by experiments describing deeper traps \cite{bemski58, levitt61,abakumov91,reshchikov14}. In the quantum picture, migration of electrons over the lattice sites is governed by a hopping matrix element, which is not explicitly temperature-dependent. The temperature influence on the capture coefficient is indirect, it enters from the effects of level broadening, thermal lattice expansion and thermal excitation of phonons. Due to highly non-linear dependence of the transition matrix element on the inter-ionic distance, the exact behavior of the capture rate with temperature is difficult to simulate quantitatively.

\section{Discussion}

We have constructed a model describing non-radiative capture and release of electrons (charge carriers) from/to the CB. The release process is divided into two stages: absorption of phonons at the trapping site, and resonant tunneling to the CB. The first stage (governed by Bose-Einstein statistics) excites the electron sufficiently in order to allow resonant escape from the trapping potential and delocalization into the continuum of the CB. The second stage sets the rate at which the localized electron leaves the trap to the CB with no gain or loss of energy. The capture process is mutually complementary to the release process, which was obtained without enforcing the detailed balance condition.

The constructed model brings out some correlating and contradicting results when compared to the semi-classical description. The classic equations are indeed obtained as limiting case of the proposed model (see Appendix~\ref{Appclaslim} for details) upon inclusion of quantum effects. Below we address these critical insights and differences.

\subsection{Coulomb neutral, active, and compact traps}\label{subsec:trapnature}

The capture coefficient (and frequency factor) calculated via the proposed model reflects the classical result in parts: there are both a velocity-like factor (units of cm~s$^{-1}$) and an effective cross section (cm$^{2}$) of the defect, see eq.~(\ref{finestrate}). In the classical approach these two factors are explained as the thermal velocity of an electron and the geometric cross-section of the trap \cite{chen11}, while in the proposed approach presented in this work, the factors consist of several host-related constants and only the effective trap radius can change from one defect to another. Furthermore, the recalculated capture coefficient has crucial additional features when compared to the classical formula. The additional dimensionless factors show a significant dependence on the wavefunction overlap of the trap states with the surrounding lattice (by means of the delocalized conduction band states), thus connecting the resonant escape of a trapped electron to a charge-transfer type of transition.

\textcolor{black}{Also the rationale behind orders of magnitude difference in capture coefficients/frequency factors for Coulomb active and neutral impurities immediately finds its explanation: the interaction matrix elements in eq.~(\ref{neutraleff}) explicitly include single particle potentials.}
Likewise, the dependence on a wavefunction overlap in the formula, which tends to decay exponentially with distance, explains the orders of magnitude variation of frequency factor between classes of compounds (ionic, covalent, (in)organic materials), and between different types of defects (impurity, structural defect) found in the same compound \cite{vedda08, Drozdowski14, Brylew14}.

The model suggests a strong dependence of the capture coefficient/frequency factor on the density of charge distribution, i.e. the covalency or ionicity of the chemical bonds within the host compound. Experimental limitations (frequency factor is obtained with $\pm$ order of magnitude accuracy) do not allow to readily quantify this effect. Instead, we connected the model to effective mass theory, which is known to show interrelation between effective mass value and covalency of the host, see Appendix~\ref{Appclaslim}.

\subsection{The connection between capture and release rates}

The dependence of the (de-)trapping process on the interaction with phonons in our proposed approach is not directly portrayed in the frequency factor. Instead, it is displayed in the Boltzmann factor (eq.~(\ref{shortrate}),~(\ref{freqfactor})). The phonon energy accumulated on the site is the potential energy of the trapped electron when the surrounding ions are in their turning positions. This configuration of the site with a trapped electron (plus a large amount of phonons creating maximum displacement) \textcolor{black}{transforms into a configuration instance of the empty trap with a new equilibrium (with no phonons left).}

The role of phonons is not only to provide additional energy to an electron, but to also reproduce the lattice configuration of a relaxed center of lower charge. The condition that the lattice is ``frozen'' in the moment of electron release allows the assumption of a single particle potential for hopping in and out of the trap. In that case, the capture ($W_{\mathrm{res\,capt}}$) and release ($W_{\mathrm{res\,rel}} \equiv s$) rates only differ by the density of states available for occupation:

\begin{equation}
     \frac{W_{\mathrm{res\,capt}}}{W_{\mathrm{res\,rel}}} \sim \frac{N_{tr}}{N_c}
     \label{roughratio}
\end{equation}

\noindent where $N_c$ is the effective density of states available in the CB, 
$N_{tr}$ is the concentration of localized states. In our approximation, the probabilities to capture or release charge carriers directly depend on the difference in effective density of the CB and the trapping states.

\subsection{Capture rate and trap depth}

In this section, we investigate if the resonant capture of CB electron is a uni-directional process (Stage~II~$\rightarrow$~Stage~I in Fig.~\ref{scheme}). In principle, immediately after resonant capture, an electron can be resonantly released as well. The resonant in-and-out hopping of the defect is only limited by its rate and the resonance condition. During capture the main energy loss process is the interaction with phonons. Significant movement of ions towards a new equilibrium geometry (due to extra charge on the lattice site) governs the rate at which the energy dissipates and the resonance is lost:
 
\begin{equation}
(E_T+E_{CB}^{kin})\frac{1+\cos(\omega_0 t)}{2}=E_T
\end{equation}

\noindent where $E_T$ is the trap depth, $E_{CB}^{kin}$ is the kinetic energy of electron ($E_{CB}^{kin}$=$k_{\mathrm{B}}T$ for thermalized electrons), $\omega_0$ is the optical phonon mode of the host. Then the time ($t_{\mathrm{off\,res}}$) (and the rate ($\alpha$)) needed for the loss of the resonance is:

\begin{equation}
\alpha = \frac{1}{t_{\mathrm{off\,res}}} = \frac{\omega_0}{2} \sqrt{\frac{E_{CB}^{kin}+E_T} {E_{CB}^{kin}}} \approx  10^{14}~\text{s}^{-1}
\label{offresest}
\end{equation}

Implementing typical values for the frequency   \textcolor{black}{effective phonon mode} $\omega_0$=3$\cdot$10$^{13}$~sec$^{-1} $\cite{lin18,bachmann09}, room temperature $k_{\mathrm{B}}T$=0.025~eV, and assuming a trap with depth $E_T$=1~eV, we obtain a rate of resonance loss due to ion motion as high as $\alpha = 10^{14}~\text{s}^{-1}$.
The obtained value for the phonon-assisted relaxation out of resonance is orders of magnitude faster than typical resonant release rate (i.e. frequency factor). Thus, once the CB-electron is resonantly captured at a sufficiently deep trap, the localization process is indeed a uni-directional process. 

\textcolor{black}{The co-factors - the effective phonon mode $\omega_0$ and the ratio of the energies - in eq.~(\ref{offresest}) can be evaluated experimentally or theoretically on a case by case basis. 
In our specific example above, the trap depth of Ce$^{3+}$ 5d-state in YAG is well-known from photo- and thermoluminescence experimentally\cite{dorenbos08}, and cryogenic high-resolution optical spectroscopy\cite{ogieglo13, bachmann09} shows the phonon modes that this center couples to (at least in the optical transition to the ground state). However, in general, the theoretical investigations of the effective phonon mode $\omega_0$ requires complex calculations of electron-phonon interaction, an-harmonic phonon couplings, and local oscillation modes, which go beyond the scope of our work. }

In the presented model, the trap depth and the capture rate are decoupled, we do not need to implement complete multi-phonon emission\cite{passler76} into Fermi's golden rule in order to describe the capture dynamics. 
\textcolor{black}{Here lies the fundamental difference in the way we treat the capture and release processes from other models \cite{abakumov85, freysoldt2014, Alkauskas14, Shi15, Turiansky21}. In order to release an electron to the CB the site needs to accumulate a huge number of phonons, while during capture only a few phonons have to be lost from the site for the electron to become trapped.} Just the loss of excess thermal energy of the thermalized CB electron ensures its localization, thus deeper traps do not necessarily have smaller capture cross sections.

\subsection{The shallowest trap}

The trap with a depth below $\delta_{E_{ph}}/2$ is not defined and should be rather perceived as an Urbach tail\cite{Urbach53} of the delocalized band. A charge carrier scatters on such a shallow defect instead of being captured. It limits the minimum possible trap depth and consequently the minimum possible time for de-trapping. In a typical oxide the frequency factor is $s=10^{11}-10^{12}$~s$^{-1}$\cite{ueda15}, while the uncertainty broadening is $2\delta_{E_{ph}}\approx$0.2~eV\cite{Alivisatos88, johns16, Bok20}. At room temperature, it is $k_{\mathrm{B}}T$=25~meV, i.e. the fastest de-trapping time ($\tau$) from the most shallow trap is

\begin{equation}
\tau = \frac{1}{s} e^{\frac{\delta_{E_{ph}}}{2k_\mathrm{B}T}}= 5\cdot 10^{-11} - 5\cdot 10^{-10}~\text{s}
\end{equation}

\noindent The shallowest trap is estimated to have a de-trapping time of at least 50-500 ps in an oxide material at room temperature. Faster processes observed in kinetics involving the delocalized bands (e.g. scintillation rise-times\cite{Derenzo00}, or ionization and relaxation rates in TA experiments\cite{Lucchini18, Tamulaitis20}) have to be connected to scattering instead of de-trapping with a Boltzmann factor. The distinction is rather important, as de-trapping is exponentially dependent on temperature, while scattering does not require any elevated temperatures to occur. This notion is supported by the independence of scintillation rise-times from temperature\cite{weele14,weele15}.

\subsection{Capture and release rates for LSO:Ce,Ca} 

Experimentally, the rate of sequent localization of an electron and a hole is encoded in the rise time of a scintillation flash. Typical values for the rise time are in units of ns\cite{Derenzo00}. Significantly faster rise times have been reported for the scintillation flash following the capture of one charge only, e.g. by the process of 

\begin{equation*}
Ce^{4+}+e^-=(Ce^{3+})^\ast \rightsquigarrow h \nu
\end{equation*}

\noindent in YAG:Ce,Mg\cite{wu14, nikl14} and LSO:Ce,Ca\cite{yang09}. In YAG:Ce~0.1\%,Mg~0.1\%\cite{gundacker16, Lucchini16, Lucchini17, gundacker18}, partial transformation of Ce$^{3+}$ to Ce$^{4+}$ gave rise to a 100~ps rise-time, while the complete shift to Ce$^{4+}$ in LSO:Ce~0.1\%,Ca~0.2\% leads to a 7~ps rise time\cite{gundacker18}. We can use this data on LSO:Ce,Ca to estimate the characteristic capture and release rates for the CB-electrons: 

\begin{eqnarray}
W_{\mathrm{res\,rel}} = A_{\mathrm{res}} \times N_c \\
W_{\mathrm{res\,capt}} = A_{\mathrm{res}} \times N_{tr}    
\end{eqnarray}

\noindent where for LSO:Ce,Ca Ce$^{4+}$ concentration is $\sim$0.1 mol\% ($N_{tr}$=10$^{19}$~cm$^{-3}$), $N_c$ is calculated from eq.~(\ref{eq:nc}), $W_{\mathrm{res\,capt}}$ is the inverse of the scintillation rise time of 7~ps. Then the capture coefficient and the frequency factor are:

\begin{eqnarray}
A_{\mathrm{res}}^{LSO} \approx 10^{-8}~\text{cm}^{3}\text{s}^{-1} \\
W_{\mathrm{res\,rel}} \equiv s_{LSO}\approx 10^{11}~\text{s}^{-1}
\end{eqnarray}

Similarly, the microscopic parameters can be estimated from the known Ce$^{4+}$ concentration and scintillation rise times for other compositions. After obtaining the capture 
coefficient and the rates from the extreme condition of only Ce$^{4+}$ present in the chosen compound, a reverse approach can be taken and the unknown Ce$^{4+}$ concentration in the given compound can be estimated from the scintillation rise time, see example on GGAG:Ce,Mg \cite{khanin19}. A test of this approach for other compounds than LSO:Ce and YAG:Ce materials is, however, still required for more accurate statements.

\subsection{Temperature dependence of the capture coefficient}

Another important consequence of the proposed model, which cannot be interpreted correctly in the classical approach, is the temperature dependence of the capture coefficient, and, more specifically, the non-vanishing value of this coefficient at low temperatures. The conventional description of electron capture requires the particle to have a defined velocity. The way to implement charge carrier velocity into a classical interpretation is via temperature, therefore a decrease of thermal velocity is expected to bring slower charge carrier capture with a $\sqrt{T}$-dependence, and to go to zero at low temperatures, see Fig.~\ref{tempdepbasic}.

The consequences of the classical approach do not agree with experiment. An efficient localization of charge carriers at LHe temperatures is regularly observed: many scintillators exhibit excellent performance at cryogenic temperatures \cite{Mikhailik10}. Furthermore, temperature-dependent studies \cite{weele14,weele15} of the scintillation time profiles show no temperature variation in their rise time between 80 and 500~K. In our approach the capture coefficient, eq.~(\ref{finestrate}), also has a velocity-dependent term, which consists of a number of constants and the trap effective radius, but the temperature dependence enters due to the effects of level broadening, thermal lattice expansion and phonon occupation within the transition matrix element.

\section{Conclusion}

A qualitative quantum mechanical description for (de)trapping of charge carriers in optical materials was established. The two-stage model is based on Fermi's golden rule, a single particle approach and thermal occupation of phonons based on Bose-Einstein statistics. The model offers a physical interpretation of the capture coefficients and frequency factors, and seamlessly includes the Coulomb-neutral/attractive nature of traps. In our interpretation the capture coefficient is not governed by thermal velocity and geometric cross-section (as the semi-classic result), instead it is very sensitive to peculiarities of the interaction potential and hybridization of the CB and localized states. The units of [cm$^{-3}$s$^{-1}$] for capture coefficient are maintained by host related basic constants.

The model offers a qualitative explanation for orders-of-magnitude variations in frequency factors observed in experimental studies, which can be related to very low overlap of wave-functions for trap and the CB states leads to low escape probabilities. Separation of the resonance condition from the accumulation/dissipation of phonons on the site have led to the supposition that the capture cross-section does not significantly depend on the trap depth (for the case of $E_T \gg k_\mathrm{B}T$). Only a fraction of phonon period is required to lose the resonance condition between the CB states and the trapping site. It can be speculated that deeper traps do not necessarily have weaker capture rates due to the substantial impact of wavefunction overlap. This is in contrast with many calculations in literature. The model suggests a strong dependence of the capture coefficient/frequency factor on the density of charge distribution, i.e. the covalency or ionicity of the chemical bonds within the host compound. Experimental limitations (frequency factor is obtained with $\pm$ order of magnitude accuracy) do not allow to readily quantify this effect. Instead, we connected the model to effective mass theory, which is known to show interrelation between effective mass value and covalency of the host.

We have also omitted the non-equilibrium process for phonon-assisted relaxation after resonant capture of electron within our model, an-harmonic phonon oscillations, electron-phonon coupling and many other effects. We believe that significant improvements to the model can be made by state-of-the-art first principle calculations correlated with the addressed scintillation kinetics and transient absorption experiments.

As such, the model generates reliable information about trap states whose exact nature is not completely understood from experimental data. 
\textcolor{black}{Overall the model connects the photo- and thermo-luminescence experiments to the nature of the wavefunctions governing traps under study.  The derived cross-sections and their evaluation with our model create feasible boundary conditions for the possible spatial extent of wavefunctions for the localized state. We show that using Shannon radii for well-known dopants, acting as traps, our model is able to reproduce experimentally derived capture coefficients. The model allows to use these experiments as one of validation tools for \textit{ab~initio} calculations, the same way as paramagnetic-resonance and Raman/IR-spectroscopy are already used\cite{Watkins98, McCluskey00}. The reverse connection has potential, where wavefunctions of localized and delocalized states carefully established by calculations are used to predict charge capture phenomena.} Explicit discussion of the physical nature of specific traps and recombination sites allows to do materials research in a more systematic way.

\begin{acknowledgements}

E. D. C. and I. I. V. acknowledge the support from the Russian Science Foundation (project 21-72-30020)
\end{acknowledgements}

\appendix

\section{Phonon statistics}\label{Appphonon}

Consider a sample that has a temperature below the Debye limit. This condition allows to neglect acoustic phonons which are not able to induce significant variations of inter-ion distances, and work only with optical phonons, which are in fact mutual oscillations of two sub-lattices including different atomic species. Now consider the energy of a specific optical mode active in a crystal. Here we do not consider phonon-phonon interactions, and ignore also a presence of  impurities/defects:

\begin{equation}
    E_{\mathrm{tot}}= \frac{V\mathrm{s}}{ (2\pi)^3 } \int_{\mathrm{BZ}}\dd[3]{k} E(k) D(k)
\end{equation}

\noindent where $V\mathrm{s}$ is the volume of the sample, $E(k)$ is the phonon dispersion relation, $D(k)$ is the average phonon occupation number. Applying simplifying assumptions that optical phonons are dispersion-less particles with the energy $\hbar \omega_0$, and adopting thermodynamic expression for equilibrium average number of bosons per atomic site at the low temperature limit ($\hbar \omega_0 > k_\mathrm{B}T$) one can get:

\begin{multline}
    E_{\mathrm{tot}}=  \frac{V\mathrm{s}}{ (2\pi)^3 } \frac{ N' \hbar \omega_0}{ \mathrm{e}^{\frac{\hbar \omega_0}{k_\mathrm{B}T}} -1 }  \int_{\mathrm{BZ}}\dd[3]{k} \approx
    \\ \approx N' \hbar \omega_0 \mathrm{e}^{-\frac{ \hbar \omega_0}{k_\mathrm{B}T}} V_s N_{uc} 
\end{multline}

\noindent where $N_\mathrm{uc}$ is the unit cells concentration, and $N'$ is the number of ions in the unit cell. This means, that on average there is an exponentially decaying probability to excite the trapped electron out of impurity site by given thermal energy $k_\mathrm{B}T$:

\begin{equation}
    \bar{E}_\mathrm{impurity} =  \hbar \omega_0 \mathrm{e}^{-\frac{\hbar \omega_0}{k_\mathrm{B}T}}
    \label{quantumenergyimpurity}
\end{equation}

On the other hand the very same results can be obtained using phonon occupation number ($n$) introduced in the following form: 

\begin{multline}
    \rho(n)=(1-\mathrm{e}^{-\frac{\hbar \omega_0}{k_\mathrm{B}T}}) \mathrm{e}^{-\frac{n \hbar \omega_0}{k_\mathrm{B}T}}, \text{~noting~that~} \\
    \sum_{n=0}^{\infty} \rho(n)=1 \text{~and~}
    \sum_{n=0}^{\infty} n \hbar \omega_0 \rho(n) =  \hbar \omega_0 \mathrm{e}^{-\frac{\hbar \omega_0}{k_\mathrm{B}T}}
    \label{probdensfun}
\end{multline}

Also, we emphasize the following feature of the probability density function: to accumulate any additional energy higher than the minimum energy of $E_{min}$ it is needed to host at least:

\begin{equation}
    n_{min} = \frac{E_{min}}{ \hbar \omega_0 } -\frac{1}{2}
\end{equation}

\noindent phonons on an impurity site. One can check that the probability to achieve this condition is:

\begin{equation}
    P( E \ge E_{min}   ) = \sum_{n=n_{min}}^{\infty} \rho(n)= const
\end{equation}

\noindent The probability to accumulate more than the minimum required number of phonons is constant for any given phonon energy $\hbar \omega_0$.

\section{Broadening for phonon occupation number}\label{Appsmearing}

\begin{figure}[]
\centering
\includegraphics[scale=0.31]{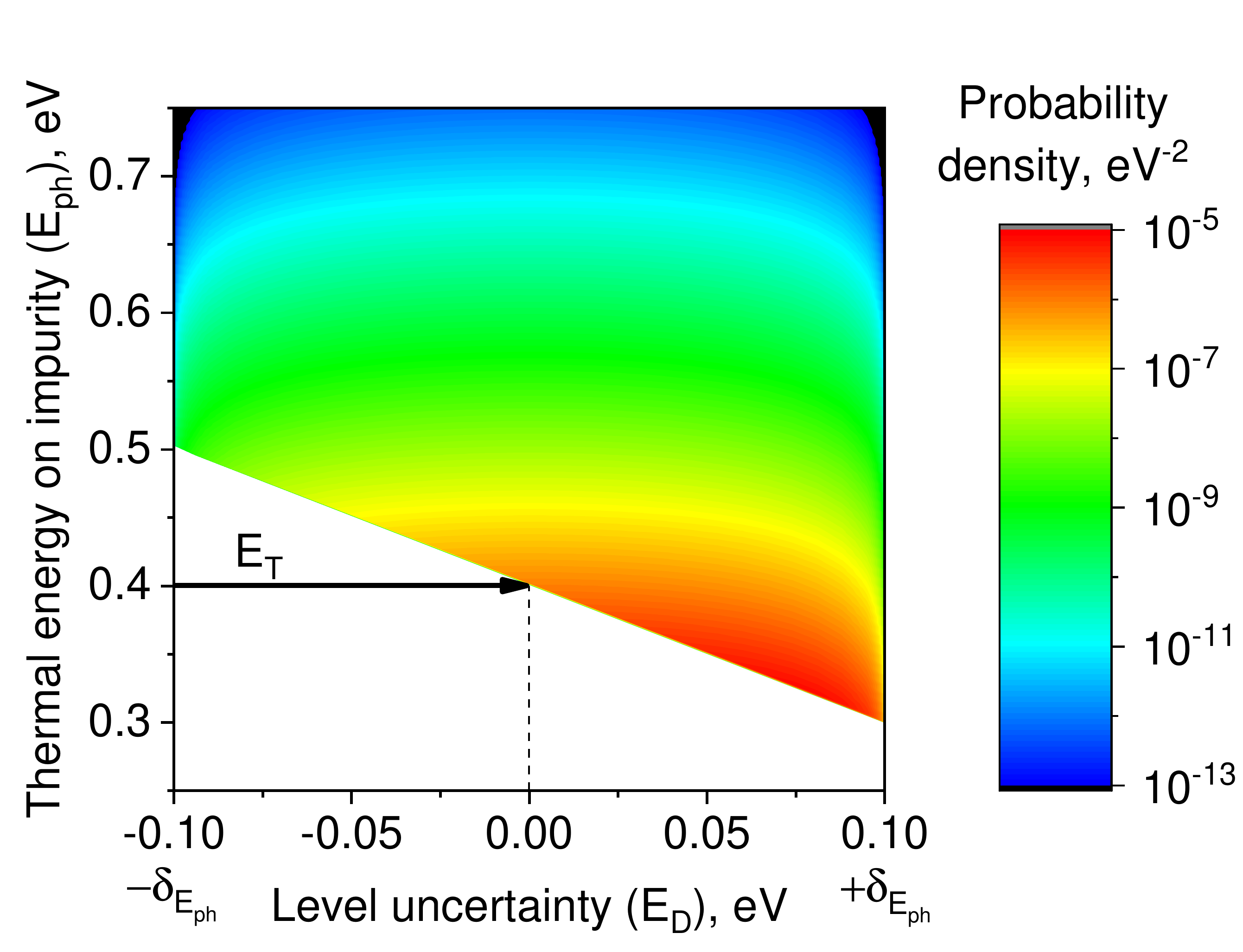}
\caption{The probability density for an electron occupying the localized level to become aligned with the CB states by interaction with phonons.}
\label{depthsmearing}
\end{figure}

For feasible numerical integration and demonstration of the general principle, we adopt a cut parabola as shape function for broadening of the localized level:

\begin{equation}
    \rho(E_\mathrm{D})=\frac{3}{4 \delta_{E_{ph}}^3}(\delta_{E_{ph}}^2 - E_D^2)
    \label{eq:broadening}
\end{equation}

Combination of the probability distribution of phonon occupation of the given localized level with the uncertainty broadening of this level allows to quantitatively assess the thermal probability for activation of the trapped electron into the CB with varying number of phonons. The trap-site  becomes levelled with the CB with probability P($E_{\mathrm{ph}}$), see Fig.~\ref{depthsmearing}.

\section{Release rate integration}\label{AppInt}

\begin{figure}[h!]
\centering
\includegraphics[scale=0.31]{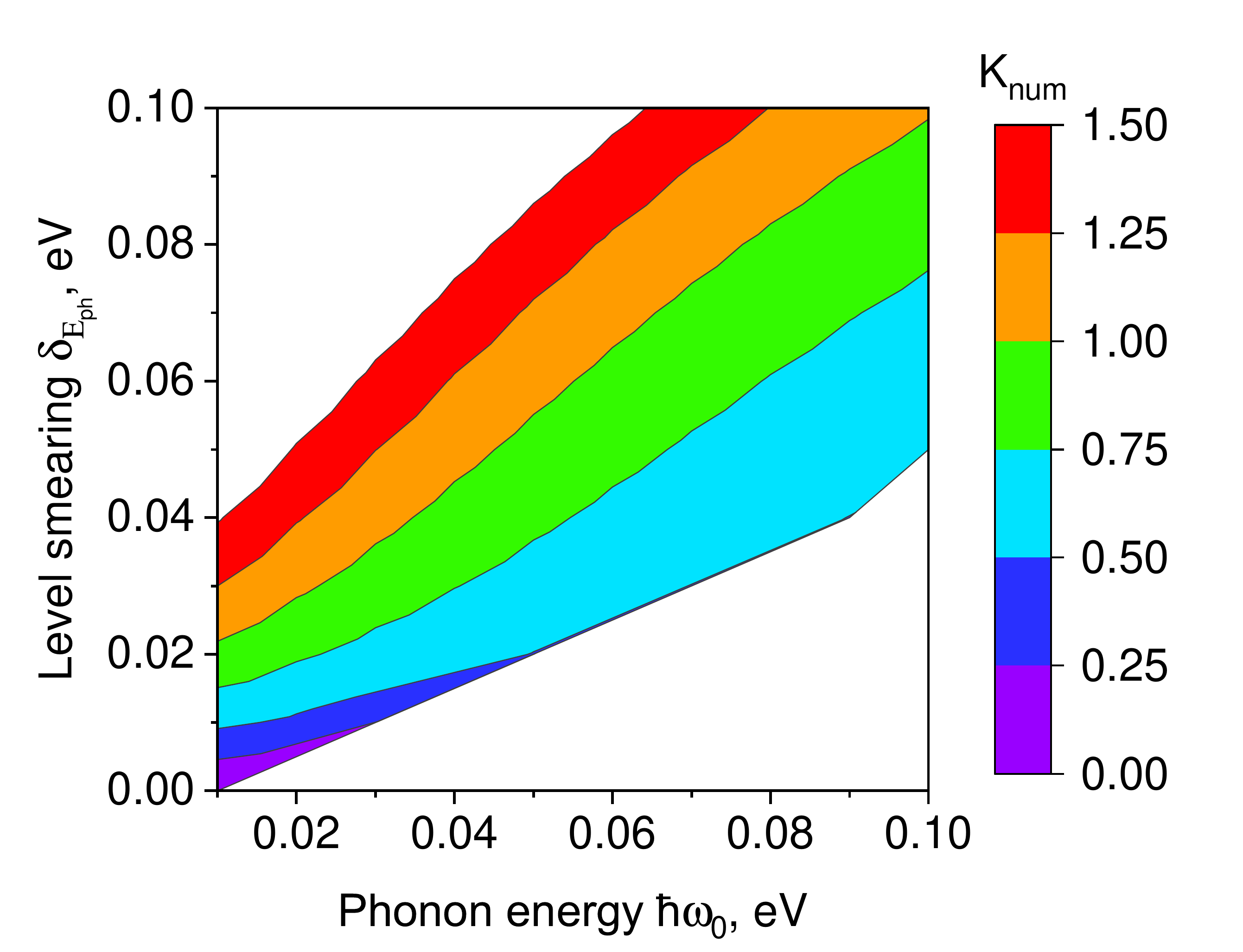}
\caption{The dependence of numerical coefficient on phonon energy and level broadening for approximation of thermal release rate.}
\label{figknum}
\end{figure}

\begin{figure*}[t]

\begin{multline}
    \int\limits_{E_{CB}^{kin} \ge 0 } \sqrt{E_{CB}^{kin}} \times dP(n,E_T) =\qty{\int\limits_{ E_T + \delta_{E_{ph}}} ^ {\infty} dE_{ph} \int\limits_{-\delta_{E_{ph}}}^{+\delta_{E_{ph}}} dE_T  +  \int\limits_{E_T - \delta_{E_{ph}}} ^ {E_T + \delta_{E_{ph}}} dE_{ph}  \int\limits_{E_T-E_{ph}}^{+\delta_{E_{ph}}} dE_T} \qty((1-e^{-\frac{\hbar \omega_0}{k_\mathrm{B}T}}) e^{-\frac{E_{\mathrm{ph}}}{k_\mathrm{B}T}} \frac{1}{\hbar \omega_0}) \times\\
    \times \qty{\frac{3}{4\delta_{E_{ph}}^3} (\delta_{E_{ph}}^2 - E_T^2)} \sqrt{E_\mathrm{ph} - E_T +E_\mathrm{D}}
\label{hugeint}
\end{multline}

\begin{multline}
    \gamma_{\mathrm{ion}} = \int  W_{res.rel.} dP =  \frac{2\pi}{\hbar} V_{\mathrm{sc}} |M_{model}|^2  \frac{8\pi}{2} \qty{\frac{2 m^\ast}{h^2}}^\frac{3}{2} K_\mathrm{num}  e^{-\frac{E_T-\frac{\delta_{E_{ph}}}{2}}{k_\mathrm{B}T}} \frac{1}{4 \delta_{E_{ph}}} \sqrt{\pi} (k_\mathrm{B}T)^{\frac{3}{2}}=\\
    = \frac{2\pi}{\hbar} V_{\mathrm{sc}} \abs{\mathcal{M}_{\mathrm{\mathrm{model}}}}^{2} \frac{1}{\delta_{E_{ph}}} K_\mathrm{num} N_c e^{-\frac{E_T-\frac{\delta_{E_{ph}}}{2}}{k_\mathrm{B}T}}
\label{resofhugeint}
\end{multline}

\end{figure*}

In this paragraph we perform the integration of eq.~(\ref{resrelgenview}) numerically, see eq.~(\ref{hugeint}). The integration in the formula (\ref{hugeint}) is excellently approximated by the following expression:

\begin{equation*}
    K_\mathrm{num}   e^{-\frac{E_T-\frac{\delta_{E_{ph}}}{2}}{k_\mathrm{B}T}} \frac{1}{4 \delta_{E_{ph}}} \int\limits_0^{\infty}e^{-\frac{E_{CB}^{kin}}{k_\mathrm{B}T}}\sqrt{E_{CB}^{kin}}~dE_{CB}^{kin}.
\end{equation*}

The  $K_\mathrm{num}$ coefficient has no physical meaning being unavoidable contribution resulting from several contributions: approximation of the broadening function, substitution of summation by integration over the phonon occupation numbers, and interpolation of the numerical integration of eq.~(\ref{resrelgenview}). The value of $K_\mathrm{num}$ depends on the localized level broadening $\delta_{E_{ph}}$ and phonon mode energy $\hbar \omega_0$, see Fig.~\ref{figknum}. Typical $K_\mathrm{num}$ value for $\delta_{E_{ph}} \approx \hbar \omega_0$ (green domain) is close to unity. 

The natural value of broadening $\delta_{E_{ph}}$ that can be adopted is $\delta_{E_{ph}} \approx \hbar \omega_0$. If it is larger than the phonon energy, then the phonon occupation number loses its meaning and must be shifted and renormalized. Otherwise, if it is significantly smaller than $\hbar \omega_0$ then the probability of thermal excitation cannot be considered as a continuous function (see eqs.~(\ref{occnumdiscr})~and~(\ref{occnumcont})).

Now, after performing numerical integration and estimating the $K_\mathrm{num}$ value, we can rewrite the resonant release rate approximation in the final form, see eq.~(\ref{resofhugeint}). Here we have arrived at eqs.~(\ref{resrelgenview}) and (\ref{resrelfinres}) of the main body of the manuscript.

\section {Effective mass limit}\label{Appclaslim}

Let us consider the fundamental correlation of the presented quantum model with the standard effective mass theory\cite{halperin60}. Below we connect them in a formal way. 

On the one hand, in the description of the CB-electron as a wave-packet its group velocity is introduced as:

\begin{eqnarray}
    \upsilon_{gr}=\frac{1}{\hbar}\frac{\partial E}{\partial k}.
\label{groupvel}
\end{eqnarray}

\noindent Assuming parabolic dispersion of a conduction band minimum the group velocity is rewritten via effective mass as:

\begin{equation}
    \upsilon_{gr}= \frac{1}{\hbar}\frac{\hbar^2 k}{m^\ast}=\frac{p}{m^\ast}
\label{effmass}
\end{equation}

\noindent where $m^\ast$ is the effective mass.

On the other hand, we can write out our capture coefficient in a more transparent way. In section~\ref{sebsec:coulomb} we have obtained the capture coefficient in the form of eq.~(\ref{finestrate}):

\begin{equation}
    A_{\mathrm{res}}=  16\sqrt{2} \pi^{\frac{3}{2}} \kappa^2 e^{-\kappa^2}\times
    \frac{1}{\frac{\delta_{E_{ph}}}{Ry^\ast}} \times
    \frac{Ry^\ast}{\hbar} a_\ast  \times  
    \pi a_\ast^2
\label{finestrate2nd}
\end{equation}

\noindent the first two terms are dimensionless coefficients describing the overlap of wavefunctions and the uncertainty broadening of states, the $\frac{Ry^\ast}{\hbar} a_\ast$ term has the units of velocity [cm~s$^{-1}$], and $\pi a_\ast^2$ has the units of geometric cross-section [cm$^{2}$]. Decomposing $Ry^\ast$ as doubled kinetic energy $Ry^\ast=\frac{p_\ast^2}{2m_e}$, $\hbar=p_\ast a_\ast$ where $p_\ast$ is the electron momentum on impurity orbital, $m_e$ is the electron mass, we obtain:

\begin{equation}
    A_{\mathrm{res}}=  D_1 \frac{1}{D_2} \frac{p_\ast}{ m_e } \pi a_\ast^2
\label{simplfinestrate}
\end{equation}

\noindent the first two terms ($D_1$, $D_2$) are repeated dimensionless coefficients, the ${p_\ast}/{ m_e }$ term has the units of velocity [cm~s$^{-1}$], and $\pi a_\ast^2$ has the units of geometric cross-sections [cm$^{2}$]. By substituting the electron mass $m_e$ by a coefficient $m_{\mathrm{eff}}$ that contains all the dimensionless terms we write $A_{\mathrm{res}}$ in a short form that is very similar to eq.~(\ref{effmass}):

\begin{eqnarray}
    A_{\mathrm{res}} = \frac{p_\ast}{ m_{\mathrm{eff}} } \pi a_\ast^2  = \upsilon_{\mathrm{eff}} \pi a_\ast^2
\label{Aeffect}
\end{eqnarray}

\noindent the introduced $\upsilon_{\mathrm{eff}}$ coefficient in our model is still the description of the CB-electron behavior, and has the same form as the eq.~(\ref{effmass}) for group velocity in the effective mass model. 

The direct correlation of the effective mass approach with our proposed model allows us to strengthen some of our discussion points. In discussion of our model (section~\ref{subsec:trapnature}) we showed that the frequency factor depends on the overlap of defect- and CB-states. We suppose that it should lead to lower (higher) frequency factors in ionic (covalent) compounds due to close (loose) packing of charge density.

Similar conclusion has been drawn for the dependence of the effective mass on the covalency/ionicity of the host\cite{rakita19}.  In the Ref.~\onlinecite{rakita19} the covalency of the host was connected to structural polarizability ($\frac{\epsilon_{\mathrm{ion}}}{\epsilon_{\mathrm{electron}}}\approx \frac{\epsilon_{s}}{\epsilon_{\infty}}-1$) of a material and bandgap-pressure coefficient ($\frac{dE_{g}}{dp}$). Both parameters showed full proportionality to the effective mass of the compound.  The effective mass systematically depends on the covalency of the material\cite{rakita19}, moreover compounds with higher covalency have lower effective mass. From eq.~(\ref{Aeffect}) we see that indeed lower effective mass leads to higher capture coefficient and frequency factor.

Please note that the effective mass model is defined for the regular lattice, thus describing interaction of CB-electrons and defect with it should be handled with caution. Local distortion of the effective mass next to irregular sites is to be expected.

\bibliography{main}

\end{document}